\documentclass[twocolumn,pra,aps,superscriptaddress,longbibliography]{revtex4-1}
\usepackage{amssymb,amsmath,amsfonts}

\usepackage[latin9]{inputenc}
\usepackage{amstext}
\usepackage{enumitem,diagbox}
\usepackage{graphicx,bm,palatino}
\usepackage[colorlinks=true,linkcolor=blue,urlcolor=blue,citecolor=blue,pdfusetitle]{hyperref}

\newcommand\ket[1]{\left|#1\right\rangle}
\newcommand\bra[1]{\left\langle #1 \right|}
\newcommand{\tr}{\operatorname{Tr}}
\newcommand{\partr}{\operatorname{tr}}
\newcommand{\var}[1]{\operatorname{Var}\left ( #1 \right)}

\newcommand{\ve}[1]{\operatorname{vec}\left ( #1 \right)}
\newcommand{\lvec}[1]{\left|#1\right\rangle \rangle}
\newcommand{\D}{\operatorname{d}\!}

\usepackage{hyperref,cleveref}
\usepackage[dvipsnames]{xcolor}
\usepackage[caption=false]{subfig}

\usepackage{times}
\usepackage{bbm}

\makeatother

\begin{document}
\title{Criticality-amplified quantum probing of a spontaneous collapse model}
\date{\today}

\author{G. Zicari}
\affiliation{Centre for Quantum Materials and Technologies, School of Mathematics and Physics, Queen's University Belfast, BT7 1NN, United Kingdom}
\email{gzicari01@qub.ac.uk}
\author{M. Carlesso}
\affiliation{Department of Physics, University of Trieste, Strada Costiera 11, 34151 Trieste, Italy}
\affiliation{Istituto Nazionale di Fisica Nucleare, Trieste Section, Via Valerio 2, 34127 Trieste, Italy}
\author{A. Trombettoni}
\affiliation{Department of Physics, University of Trieste, Strada Costiera 11, 34151 Trieste, Italy}
\affiliation{CNR-IOM DEMOCRITOS Simulation Center, Via Bonomea 265, 34136 Trieste, Italy}
\author{M. Paternostro}
\affiliation{Quantum Theory group, Dipartimento di Fisica e Chimica - Emilio Segr\`e, Universit\`a degli Studi di Palermo, via Archirafi 36, I-90123 Palermo, Italy}
\affiliation{Centre for Quantum Materials and Technologies, School of Mathematics and Physics, Queen's University Belfast, BT7 1NN, United Kingdom}

\begin{abstract}

Spontaneous collapse models, which are phenomenological mechanisms introduced and designed to account for dynamical wavepacket reduction, are attracting a growing interest from the community interested in the characterisation of the quantum-to-classical transition. Here, we introduce a {\it quantum-probing} approach to the quest of deriving metrological upper bounds on the free parameters of such empirical models. To illustrate our approach, we consider an extended quantum Ising chain whose elements are -- either individually or collectively -- affected by a mechanism responsible for spontaneous collapse. We explore configurations involving out-of-equilibrium states of the chain, which allows us to infer information about the collapse mechanism before it is completely scrambled from the state of the system. Moreover, we investigate potential amplification effects on the probing performance based on the exploitation of quantum criticality.

\end{abstract}

\maketitle

\section{Introduction}
\label{sec:intro}

The quantum measurement problem, which was made famous by the Schr\"odinger-cat paradox, is one of the most controversial aspects in the current formulation of quantum theory. In order to tackle it, models for the  potential spontaneous collapse of the wavefunction of a quantum system have been built, leveraging on suitable  modifications of the quantum axioms~\cite{Bassi_rev:2003}. Specifically, the Schr\"odinger equation describing the evolution of the state of a quantum system is modified so as to include stochastic and non-linear terms. 

Consequently, the dynamics for the density operator $\hat \rho$ includes a term describing the spontaneous, fundamental decoherence-like mechanisms, which suppresses the quantum coherence in a specific basis. Among the various formulations proposed to account for this, the Continuous Spontaneous Localisation (CSL) model~\cite{ghirardi1990markov} and the Di\'osi-Penrose (DP) one~\cite{diosi1989models,penrose1996gravity} are two of the most celebrated. While the latter attributes the origin of the collapse mechanism to gravity, the former is a fully phenomenological model and it is typically taken as the paradigm of collapse models. In the last decade, the experimental verification of collapse models has attracted a strong interest~\cite{bassi2013models,carlesso2022present}. Owing to such endeavours, some important quantitative bounds on the various parameters characterising such models have been established. Importantly, the full coverage of the models' parameter space is still in need of significant advances, including the deployment of new and sophisticated techniques for their extensive exploration~\cite{Marchese1,Marchese2}.

The framework of {\it quantum probing} makes use of simple quantum systems coupled with complex environments to precisely extract information about environmental parameters, such as its temperature or spectral density~\cite{Smirne:2013,Benedetti2018,Tamascelli2020,Mehboudi2019}. The expectation values of the system observables are indeed sensitive to changes in those parameters~\cite{Mascherpa:2017}, as one can assess even in the linear response regime~\cite{Konopik:2019,Blair:2024}. 
Quantum probes are recognised for their capacity to achieve superior precision compared to classical sensors using equivalent resources~\cite{Degen2017}. Significant advantages are known to stem from the preparation of the probes in specific non-classical, yet fragile~\cite{Demkowicz2012}, states~\cite{Leibfried2004,Giovannetti2004}. A way around the problem embodied by the fragility of such an approach is offered by the exploitation of sensitivity-enhancing mechanisms based on many-body effects: the strong, long-range correlations emerging in the state of a quantum many-body system near its critical point is expected to generally enhance the sensing capabilities and result, in turn, in a higher estimation precision~\cite{CamposVenuti:2007,Zanardi:2008,Frerot:2018,Rams:2018,Hotter:2024,Mihailescu:2024}. More recently, several protocols have been introduced to show that such an enhancement is not only featured in the thermodynamic limit, but also in finite-component quantum phase transitions~\cite{Garbe:2020,Garbe:2022,DiCandia:2023,Alushi:2024}.

In this paper, we leverage criticality to {\it amplify} the effects of the collapse of a quantum many-body system near a critical point to demonstrate the criticality-enhanced estimate of the corresponding collapse rate. To this end, it is worth mentioning that collapse mechanisms can be embedded in the dynamical equations, which, after performing a suitable stochastic average, can be recast in the form of a master equation in the Lindblad form~\cite{Bassi_rev:2003,Vacchini_2007}. Within the framework offered by the theory of open quantum systems~\cite{Breuer-Petruccione,Vacchini:2024}, the collapse mechanism can be seen as a fundamental environmental effect that affects the system dynamics. Our study is \textit{not} about ascertaining how such a noise affects the ability to discriminate a given Hamiltonian parameter~\cite{Haase:2016}; rather, is about estimating the parameter that controls the noise, i.e.~the collapse rate. In other words, we tackle a metrological problem of estimating a non-Hamiltonian parameter, i.e.~entering in the dissipative part of the master equation~\cite{Fujiwara:2001,Monras:2007}. Motivated by a problem of fundamental physics, we resort to a dissipative version of the quantum Ising model as a testbed of our approach. The rationale behind this choice is that, in our proof of principle, we can simplify the spontaneous localisation problem by encoding the sudden collapse in space into two discrete degrees of freedom for each site of the Ising chain, extending to a many-body system an idea already explored in the single-particle scenario~\cite{Ghirardi:1999,Bassi:2004,Bahrami:2013}.
The problem is further complicated by the introduction of a fully controllable probe, given by a two-level system, that senses the effect of the collapse.  In this picture, the collapse mechanism acts as a perturbation on the closed system dynamics, thus we are still able to leverage the critical properties of such prototypical model of statistical mechanics~\cite{Dziamarga:2005}, especially in relation to the scopes of quantum metrology~\cite{Invernizzi:2008}. In order to investigate the influence of criticality, we explore the metrological performance of the probing system -- quantitatively provided by the Quantum Fisher Information (QFI) -- across an ample region of the phase diagram of the chain, highlighting the situations that correspond to an enhanced sensitivity of the QFI to the collapse rate. Moreover, we show that possible spatial correlations across the collapse mechanisms affecting the elements of the chain do not alter significantly the picture gathered by assuming individual events, which can thus be considered as representative of the typical performance of the proposed probing scheme.

The remainder of this manuscript is organized as follows. In Sec.~\ref{sec:probing} we give a short outline of the probing mechanism at hand, discussing in particular the use of non-equilibrium states. Sec.~\ref{sec:model} is dedicated to the description of the physical model at hand, the transverse Ising model, which is then used to identify working points where a criticality-enhanced estimate of the rate of a spontaneous collapse model is possible [cf. Sec.~\ref{sec:many-body}]. In Sec.~\ref{sec:results} we give an account of the set of results stemming from our study, and present an assessment of the effects that spatially correlated collapse mechanisms would have on the performance of our quantum probing approach. Finally, while Sec.~\ref{sec:conclusions} is devoted to both our conclusions and the forward look  stemming from our analysis, a set of  Appendices report some technical aspects of our study.

\section{Non-equilibrium probing}
\label{sec:probing}

Let us consider a typical scenario for an estimation problem. Suppose that one has a generic quantum system $S$, whose dynamics depends on a parameter $\lambda$ to be estimated. One can follow two approaches, which we can dub as \emph{equilibrium} or \emph{non-equilibrium} probing. The former corresponds to the following scenario: the dynamics of the system $S$ converges to a steady state that encodes information about $\lambda$, thus we use that state to estimate the parameter itself. By contrast, the latter non-equilibrium approach consists in letting the system $S$ interact with a second one, called \emph{probe}, that can be fully controlled. This allows us to dynamically infer information about the parameter $\lambda$ as the evolution time goes by, without directly interfere with $S$. These approaches are often used in quantum estimation problems, where they are sometimes referred to as \emph{static} or \emph{dynamical} encoding, respectively~\cite{Seveso:2020}. For instance, they are used in the context of quantum thermometry~\cite{DePasquale:2018,DeffnerCampbellBook}, where the parameter to be estimated is the system temperature~\cite{Correa:2015,Mok:2021,Campbell:2018,Brunelli:2011,Brunelli:2012}.

In a similar spirit, in what follows we will choose to apply a non-equilibrium probing scheme to a particular problem of fundamental physics. To this end, we can choose a two-level system (qubit) to serve as a probe for the system $S$. The non-equilibrium scheme is described by the following Hamiltonian: 
\begin{equation}
\label{eq:non_eq_H_scheme}
\hat{H} = \hat{H}_{S} \otimes \mathbbm{1}_p + \mathbbm{1}_S \otimes \hat{H}_{p} + \hat{H}_I \, ,
\end{equation}
where $\hat{H}_{S}$ and $\hat{H}_{p}$ are the Hamiltonian operators of the system and the probe, respectively. The Hamiltonian $\hat{H}_I$ accounts for the interaction between them, while $\mathbbm{1}_S$ and $\mathbbm{1}_p$ are the identity operators defined over the Hilbert spaces associated with the system and the probe, respectively.

This non-equilibrium scheme requires solving the dynamics of such composite system. Let us assume that the dynamics is governed by the following dynamical equation:
\begin{equation}
\label{eq:dyn_eq}
\frac{\D \hat{\rho}}{\D t} = \mathcal{L}_\lambda \hat{\rho},
\end{equation}
where $\mathcal{L}_\lambda$ is the Liouvillian (super)-operator encompassing both the unitary and non-unitary contributions. Note that $\mathcal{L}_\lambda$ implicitly depends on $\lambda$, which is the parameter we want to estimate.
 
In the following, we narrow down our selection of Liouvillians by considering those of the standard form~\cite{Breuer-Petruccione}
\begin{align}
\label{eq:liovillian_general}
\mathcal{L}_\lambda \hat{\rho} \equiv - i [\hat{H}, \hat{\rho}] + D(\hat{\rho}) \, ,
\end{align}
where $D(\hat{\rho})$ is the dissipator in the standard Lindblad form~\cite{Lindblad:1976,Gorini:1976,Gorini:1978}. Note that we will work in units such that $\hbar = k_{\text{B}}=1$. In principle, either the unitary or the non-unitary part can bear dependence on $\lambda$. In our study, however, we consider the case in which $\lambda$ is a parameter characterising the dissipator. 

Given the initial state $\hat{\rho}(0)$, one can formally solve Eq.~\eqref{eq:dyn_eq}, yielding $\hat{\rho}(t) = e^{t \mathcal{L}_\lambda} \hat{\rho}(0)$ at any time $t$. One can thus trace out the degrees of freedom of the system $S$, obtaining the state of the probe $\hat{\rho}_p(t) = \partr_S \hat{\rho}(t)$, which -- to ease the notation -- will be denoted as $\hat{\rho}_\lambda$, as the probe density matrix implicitly depends on the parameter $\lambda$ to be estimated. 

Our problem can be rephrased through the language of quantum metrology~\cite{Paris:2009,Pezze2018,Braun2018,Sidhu:2020}, where the key quantity is represented by the Quantum Fisher Information (QFI), defined as
\begin{equation}
\label{eq:QFI_def}
G(\lambda) \equiv \tr \left [ \hat{\varrho}_\lambda \hat{\Lambda}_\lambda^2 \right ] \, ,
\end{equation}
where the symmetric logarithmic derivative 
$\hat{\Lambda}_\lambda$ (such that $\hat{\Lambda}_\lambda^\dagger = \hat{\Lambda}_\lambda$) is implicitly defined through the equation
\begin{equation}
\label{eq:SLD_def}
{\partial_\lambda \hat{\varrho}_\lambda} = \frac{1}{2}(\hat{\Lambda}_\lambda \hat{\varrho}_\lambda + \hat{\varrho}_\lambda \hat{\Lambda}_\lambda) \, .
\end{equation}
The QFI provides the ultimate limit to the precision of the estimate of the unknown parameter $\lambda$, as it enters in the so-called \emph{quantum Cram\'{e}r-Rao bound} for the variance $\var{\lambda}$ of the unbiased estimator, i.e.
\begin{equation}
\label{eq:Quantum_Cramer-Rao}
\var{\lambda} \ge [M G(\lambda)]^{-1} \, ,
\end{equation}
where $M$ is the number of independent measurements performed on the quantum system. Note that the QFI is independent of the particular measurement that has been performed, thus it is an intrinsic feature of the family of the probing states~\cite{Paris:2009}. From Eq.~\eqref{eq:Quantum_Cramer-Rao}, one can easily deduce that for a given $M$, maximising the QFI $G(\lambda)$ leads to the minimum lower bound for the variance. Our aim is to investigate under which physical conditions one can maximise $G(\lambda)$.

In order to calculate the QFI, we need a closed formula which specialises the general definition in Eq.~\eqref{eq:QFI_def} to the case of a qubit. Let us suppose that we can diagonalise the density matrix $\hat{\varrho}_\lambda$
as $
\hat{\varrho}_\lambda = \varrho_{+} \ket{\psi_{+}} \!  \bra{\psi_{+}} + \varrho_{-} \ket{\psi_{-}} \!  \bra{\psi_{-}} $, which allows us to express the QFI as
\begin{equation}
\begin{aligned}
\label{eq:qubit_QFI_eig_split}
G(\lambda) &=  \sum_{j=\pm}\frac{(\partial_\lambda \varrho_{j})^2}{\varrho_{j}} 
{+} 2 \kappa \left [| \langle \psi_{-} | \partial_\lambda \psi_{+} \rangle |^2 {+} | \langle \psi_{+} | \partial_\lambda \psi_{-} \rangle |^2  \right ] \, ,
\end{aligned}
\end{equation}
with 
\begin{equation}
\kappa = \frac{(\varrho_{+} - \varrho_{-})^2}{\varrho_{+} + \varrho_{-}} = \left (1 - 2 \varrho_{+} \right )^2 \, ,
\end{equation}
where we used the normalization condition $\varrho_{+} + \varrho_{-} = 1$. The derivatives of the eigenvectors are decomposed over the basis $\{\ket{0},\ket{1}\}$ of eigenstates of the $z$ Pauli operator $\hat{\sigma}^z  = \ket{1}\!\bra{1} - \ket{0}\!\bra{0}$, i.e.
\begin{align}
\ket{\partial_\lambda \psi_{\pm}} =  \partial_\lambda{\langle 0 | \psi_{\pm} \rangle} \ket{0} +\partial_\lambda{\langle 1 | \psi_{\pm} \rangle} \ket{1}.
\end{align}

\section{The model}
\label{sec:model}

To introduce the model, we first focus on the system $S$ alone. In general, the effect of the spontaneous collapse can be embedded into the dynamical equations by including non-linear and stochastic terms~\cite{bassi2013models,Bassi_rev:2003,carlesso2022present}. A simple, yet insightful, instance is represented by a two-level system (qubit), where we associate the states $\ket{\pm} = (\ket{0} \pm \ket{1})/\sqrt{2}$ to two different spatial configurations. In this case, the stochastic and non-linear Schr\"odinger equation reads as~\cite{Bassi:2004,Bahrami:2013}
\begin{align}
\label{eq:1qubit_schro}
\D \ket{\psi_t} = \Big [ - i \hat{H}_S \D t + &\sqrt{\lambda} \left ( \hat{\sigma}^z - \langle \hat{\sigma}^z \rangle\right ) \D W_t \nonumber \\
& - \frac{\lambda}{2} \left ( \hat{\sigma}^z - \langle \hat{\sigma}^z \rangle\right )^2 \D t \Big ] \ket{\psi_t} \, ,
\end{align}
with $\hat{H}_S = (\omega_0/2) \hat{\sigma}^x$, where $\omega_0$ is the energy splitting between the two levels, and $\hat{\sigma}^x = \ket{1}\!\bra{0} + \ket{0}\!\bra{1}$ is the $x$ Pauli operator. The last two terms in Eq.~\eqref{eq:1qubit_schro} contain the constant $\lambda$, known as \textit{collapse rate}; those terms are responsible for the collapse of the system wavefunction in one of two states $\ket{\pm}$ consistently with the Born probability rule. Note that $W_t$ is the standard Wiener process~\cite{Gardiner:2009}.

Since we do not track the system dynamics at the level of single trajectories, we can rearrange Eq.~\eqref{eq:1qubit_schro} to write down the dynamical equations in terms of suitable averages. To this end, we define the statistical operator as $\hat{\rho}_S = \mathbb{E} [ \ket{\psi_t} \! \bra{\psi_t}]$, where $\mathbb{E}[\cdot]$ denotes the stochastic average. This observation allows us to recast Eq.~\eqref{eq:1qubit_schro} in the form of a master equation for the statistical operator $\hat{\rho}_S$, i.e. 
\begin{equation}
\label{eq:1qubit_ME}
\frac{\D \hat{\rho}_S}{\D t} = - i [\hat{H}_S, \hat{\rho}_S] + D(\hat{\rho}_S) \, .
\end{equation}
In this picture, the unitary part of the dynamics is controlled by the Hamiltonian $\hat{H}_S$, while the collapse mechanism is effectively seen as the effect of an external bath with which the system interacts. The latter leads to a non-unitary evolution, which can be expressed through the following dissipator in the Lindblad form~\cite{Bassi:2004,Bahrami:2013}:
\begin{equation}
\label{eq:qubit_collapse_dissipator}
D(\hat{\rho}_S) = - \frac{\lambda}{2} \, [\hat{\sigma}^z, [ \hat{\sigma}^z,\hat{\rho}_S] ] \, ,
\end{equation}
which is proportional to the collapse rate $\lambda$, i.e. the parameter we aim at estimating. Note that free Hamiltonian $\hat{H}_S$ does not commute with the operator $\hat{\sigma}^z$ entering in Eq.~\eqref{eq:qubit_collapse_dissipator}, which results in non-trivial dynamics, bringing about both decoherence and dissipation. The dynamics governed by Eq.~\eqref{eq:1qubit_ME} can be analytically solved, as shown in Ref.~\cite{Bassi:2004}. 
 
The solution is found by vectorising Eq.~\eqref{eq:1qubit_ME}, i.e.~by following the prescriptions given in Appendix~\ref{app:vectorisation}. By so doing, one can obtain the state of the system at any time $t$, i.e. $\hat{\rho}_S = \hat{\rho}_S(t)$. We can recall that the density matrix of a qubit can be written as 
\begin{align}
\hat{\rho}_S = \frac{\mathbbm{1} + \boldsymbol{\sigma} \cdot \boldsymbol{\tau}}{2} \, ,
\end{align}
where $\boldsymbol{\sigma} \equiv (\hat{\sigma}^x, \hat{\sigma}^y, \hat{\sigma}^z)$ is the Pauli vector, while $\boldsymbol{\tau} \equiv (\tau_x, \tau_y, \tau_z)$ is the Bloch vector, whose components are defined as
$\tau_k \equiv \tr(\hat{\sigma}^k \hat{\rho}_S)$, with $k=x,y,z$. 

\begin{figure}[t!]
\centering
     \begin{minipage}[t!]{0.29\textwidth}
        \centering
         \includegraphics[width=\columnwidth]{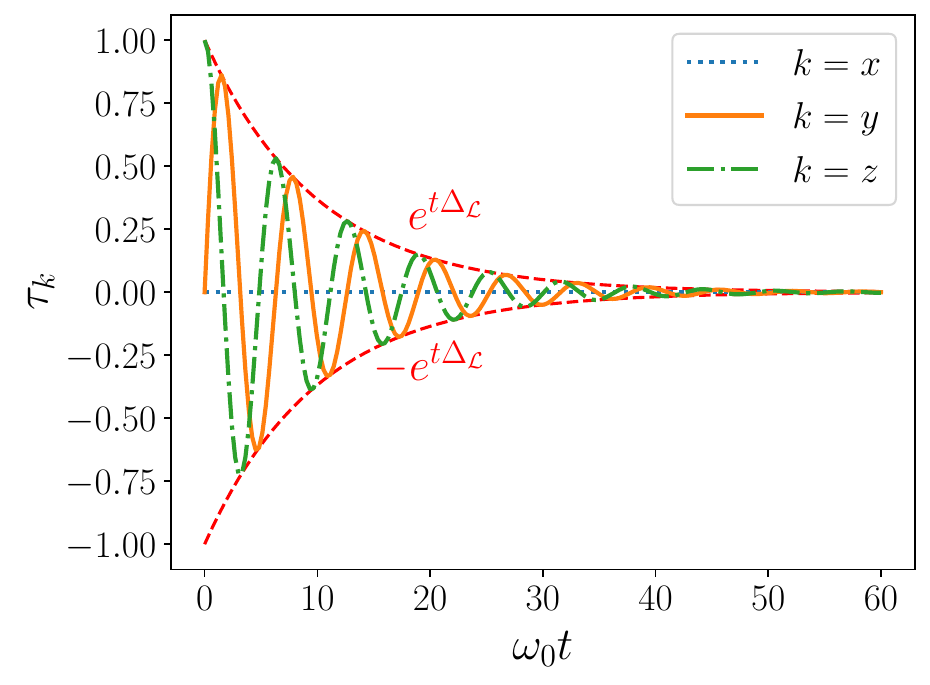}
    \end{minipage}
   \hfill
   \begin{minipage}[t!]{0.18\textwidth}
        \centering
         \includegraphics[width=\columnwidth]{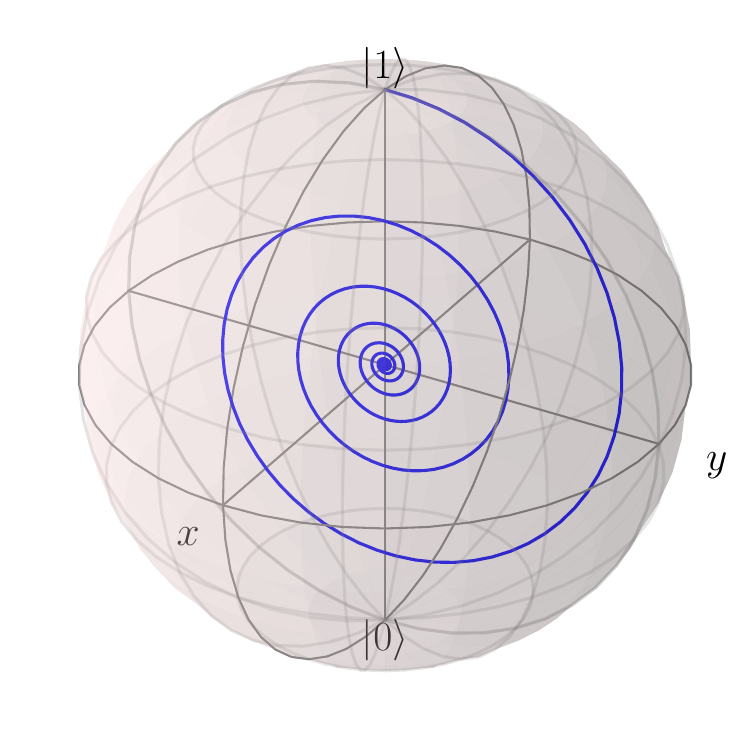}
    \end{minipage}
        \caption{Dynamics of a qubit undergoing the collapse mechanism described by the dissipator in Eq.~\eqref{eq:qubit_collapse_dissipator}. We simulate the system dynamics choosing $\ket{1}$ as initial state, and $\ \lambda=0.1\omega_0$. On the left, we show the plots of the Bloch vector components  
        $\tau_k = \langle \hat{\sigma}^k \rangle$ (with $k=x,y,x$), along with the envelopes $e^{\pm t \Delta_\mathcal{L}}$, which captures the typical relaxation time, with $\Delta_\mathcal{L}$ being the Liouvillian gap. On the right, we display the corresponding trajectory in the Bloch sphere. }
        \label{fig:Qubit_collapse_dynamics_bloch}
\end{figure}

In Fig.~\ref{fig:Qubit_collapse_dynamics_bloch}, we plot the components of the Bloch vector against time, i.e.~$\tau_k = \tau_k(t)$, together with the corresponding Bloch sphere representation. Furthermore, it is worth mentioning that one can compute the Liouvillian gap $\Delta_\mathcal{L}$, which identifies the rate of the slowest relaxation mode~\cite{Spohn:1976,Chruscinski:2022}, thus correctly providing an estimate of the typical system relaxation time.
The details of the calculation are given in Appendix~\ref{app:sol_ME}. 
One sees that the generic state at any time $t$, i.e.~$\hat{\rho}_S = \hat{\rho}_S(t)$, bears dependence on the collapse rate $\lambda$. However, this dependence is completely washed out by the dynamical process, as one can immediately deduce from the form of the steady state $\hat{\rho}_{\rm ss} =\mathbbm{1}/2$. In other words, regardless of the choice of the initial state, the system relaxes towards a unique maximally mixed state~\cite{Ghirardi:1999,Bassi:2004}. These features have an important consequence for our estimation problem: we cannot indeed infer any information about $\lambda$ by solely looking at the steady state. Therefore, in order to estimate the collapse rate $\lambda$, we need to devise a different strategy that is able to capture information about the collapse rate before the system reaches its unique steady state.

\subsection{Non-equilibrium probing}

To probe the collapse dynamics of a single qubit undergoing the spontaneous collapse mechanism described by the dissipator in Eq.~\eqref{eq:1qubit_ME}, one can follow the non-equilibrium scheme described in Sec.~\ref{sec:probing}. We thus introduce a second qubit that serves as a probe for the first one. The total dynamics is then given by  Eq.~\eqref{eq:non_eq_H_scheme}, where we take $\hat{H}_p = \omega_p \hat{\sigma}^x_p / 2$ as the Hamiltonian of the probe, while we engineer the interaction between the two qubits to be along $z$, i.e.
\begin{equation}
\hat{H}_I = g \, \hat{\sigma}^z_S \otimes \hat{\sigma}^z_p \, ,
\end{equation}
where $g$ is the coupling constant. Our scheme is based on the assumption to have full control on the probe, whereas the system is subject to the collapse mechanism, as described by Eq.~\eqref{eq:1qubit_ME}, where the Pauli operator need to be suitably defined as $\hat{\sigma}^z \to \hat{\sigma}^z_S \otimes \mathbbm{1}_p$.
Similarly to the previous case, we can solve the dynamics by vectorisation of the corresponding master equation for a given initial state [cf.~Appendix~\ref{app:vectorisation}]. Specifically, we assume that the system is initially prepared in the product state
\begin{equation}
\label{eq:qubit_initial_state}
\hat{\rho}(0) = \hat{\rho}_S(0) \otimes \hat{\rho}_p(0) \, ,
\end{equation}
where the system is in a thermal Gibbs state $\hat{\rho}_S(0) = \exp(-\beta \hat{H}_S)/\mathcal{Z}_S$, with $\mathcal{Z}_S = \tr{[ \exp(-\beta \hat{H}_S) ]}$, $\beta = 1/T$ being the inverse temperature. The probe is prepared in a pure state in the form $\hat\rho_p = \ket{\psi_p} \! \bra{\psi_p}$, where we write as usual $\ket{\psi_p} = \cos(\theta/2) \ket{1} + e^{i \phi} \sin(\theta/2) \ket{0}$, with the constraints $\theta\in[0,\pi]$ and $\phi\in[0, 2 \pi]$.

The explicit solution of the dynamics yields the density matrix $\hat{\rho} = \hat{\rho}(t)$. Thus, by tracing out the system, we can obtain the reduced state of the probe at any time $t$, i.e. $\hat{\rho}_p(t) \equiv \partr_S \hat{\rho}(t)$. 
It is straightforward to assess the qualitative effect of the collapse mechanism onto the system dynamics: larger rates $\lambda$ results in damping the oscillations over a shorter timescale. This complicates the detection of the effects of collapse for small values of $\lambda$.

\subsection{Numerical results}

Since $\hat{\rho}_p(t)$ implicitly depends on the parameter $\lambda$, and such a dependence can be used to compute the QFI via Eq.~\eqref{eq:qubit_QFI_eig_split}. It is not redundant to stress that the QFI, apart from being a function of the collapse rate $\lambda$, depends on several additional parameters: the evolution time $t$ and the angles that specify the initial (pure) state of the probe, i.e. $\theta$ and $\phi$. We verified that the function $G$ is independent of the choice of the azimuthal angle $\phi$. Differently, the role played by the parameters $t$ and $\theta$ is nontrivial. Specifically, the evolution time, depending on the magnitude of the collapse rate $\lambda$, should satisfy two physical constraints. On one hand, it should not be too small so that we depart form fully coherent oscillations; on the other hand, it should not be too large to guarantee that the amplitude of the oscillations is still non-vanishing. 

Therefore, before studying $G$ as a function of the parameter $\lambda$, we perform an optimisation over the pair $(t, \theta)$ yielding $(t_{\rm opt}, \theta_{\rm opt})$ such that $G_{\rm max} = G(t_{\rm opt}, \theta_{\rm opt})$. For instance, we can consider the situation where $\lambda = 0.1 \omega_0$, while we choose the remaining physical parameters as $\omega_p = 0.3 \omega_0$, $ g = 0.2 \omega_0$, $ \beta = 0.01/\omega_0$, and $ \phi = \pi/4$. Using the Nelder-Mead algorithm~\cite{Gao:2012}, we obtain $\omega_0 t_{\rm opt} \approx 56.82$ and $\theta_{\rm opt} \approx 3.13$. This procedure allows us to determine an optimal working point. 

\begin{figure}
	\centering
        \includegraphics[width=0.9\columnwidth]{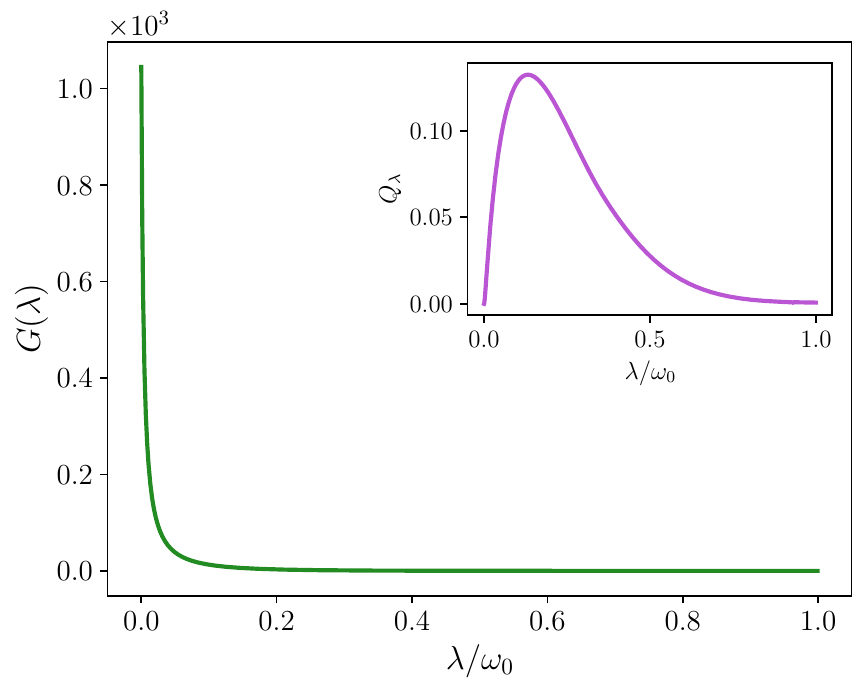}
\caption{Non-equilibrium probing of the collapse rate $\lambda$ appearing in Eq.~\eqref{eq:qubit_collapse_dissipator}. We plot the QFI $G$ as a function of $\lambda$, while in the inset we show the QSNR $Q_\lambda$. For the numerical simulations we chose $\omega_p=0.3 \omega_0$ and $g=0.2 \omega_0$. The initial state of the probe is fully specified by the angles $\theta = \phi = \pi/4$, while the system is in a thermal Gibbs state such that $\beta=0.01/\omega_0$. The maximum evolution time is given by $\omega_0 \bar{t} = 2 \pi/ g$. }
\label{fig:qubit_collapse_QFI}
\end{figure}

Alternatively, one can choose to fix $t$ and $\theta$ as $t=\bar{t}$ and $\theta = \bar{\theta}$, which essentially means to fix the maximum evolution time and the initial state of the probe. We can then study how the QFI varies as a function of the parameter we would like to estimate, i.e.~$\lambda$. Beside the QFI $G(\lambda)$, one can also look at the Quantum Signal-to-Noise Ratio (QSNR), defined as $Q_\lambda \equiv \lambda^2 G(\lambda)$, which quantifies how the variance scales with the mean value~\cite{Paris:2009}. In Fig.~\ref{fig:qubit_collapse_QFI}, we show that the QFI $G(\lambda)$ {\it monotonically} decreases with the parameter $\lambda$: the maximum precision is asymptotically achieved in the limit $\lambda \to 0$. However, we need to carefully interpret this behaviour, as the QFI indicates as regions of high precision also those where the signal itself is vanishing small, i.e. challenging to detect. Therefore, a faithful measure of the precision we can achieve is given by the QSNR, where we need to compare the information we gather through the QFI with the variance of the estimator. The inset in Fig.~\ref{fig:qubit_collapse_QFI} shows that  the best precision is obtained for a finite value of $\lambda$, around which the curve corresponding to $Q_\lambda$ is peaked.

\section{Many-body approach to the estimation of the collapse rate}
\label{sec:many-body}

\begin{figure}
	\centering
        \includegraphics[width=0.9\columnwidth]{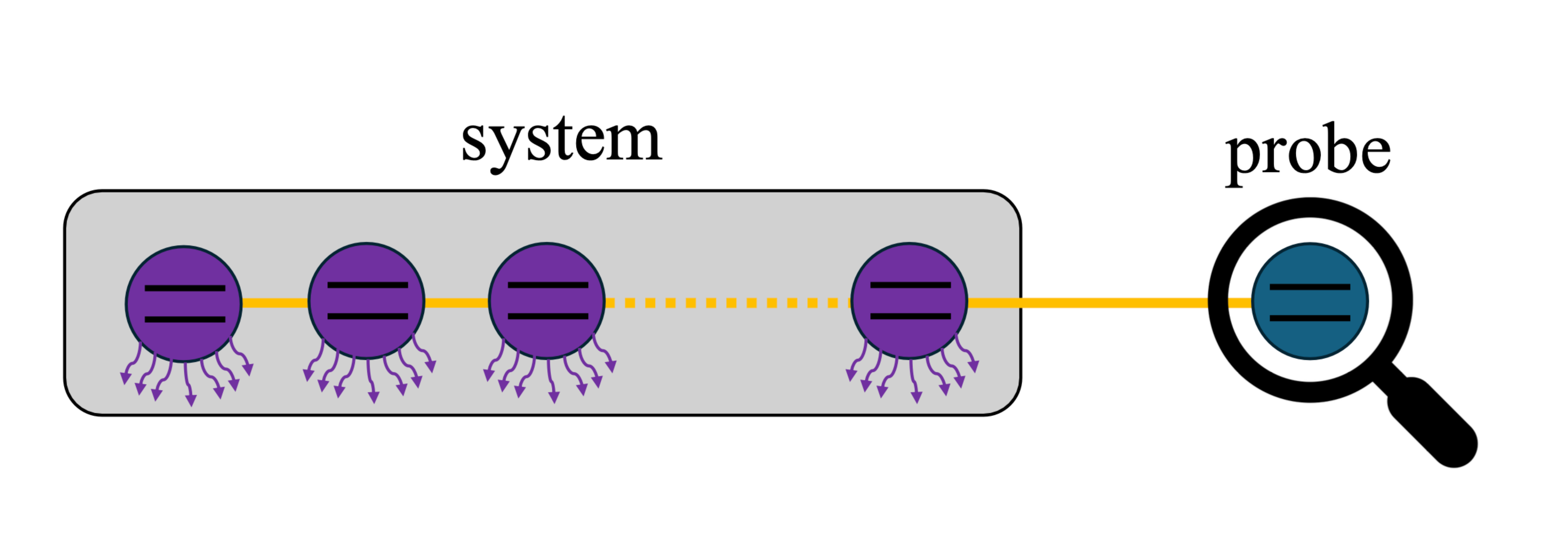}
\caption{Sketch of the non-equilibrium probing scheme to estimate the collapse rate $\lambda$. The system is made of a chain of qubits interacting in a Ising-like form: each of them is independently undergoing the collapse mechanism. The chain is attached to a perfectly isolated qubit that serves as a probe.}
\label{fig:sketch}
\end{figure}

In this Section we introduce a many-body approach to provide the best estimate of the collapse rate $\lambda$. To this end, {motivated by the discussion of the previous section}, we introduce a non-equilibrium setup. We use a single qubit to probe a many-body system affected by the collapse mechanism, as sketched in Fig.~\ref{fig:sketch}.
Our aim is essentially twofold. On the one hand, we investigate whether and under which conditions quantum criticality~\cite{Sachdev:2011,Carollo:2020} is resourceful for our estimation problem. On the other hand, we assess how the estimability of the parameter $\lambda$ scales with the system size $N$, with a focus on small values of $N$. The former hinges on a key result of quantum metrology: quantum criticality represents a resource for the estimation of Hamiltonian parameters~\cite{CamposVenuti:2007,Zanardi:2008}. The latter, instead, can also be regarded as a way to assess the existence of cooperative effects in a many-body system, which could enhance our ability to estimate a single parameter.

To address the aforementioned issues, we consider the prototypical example of a Transverse-Field Ising Model (TFIM). The Hamiltonian of the system reads as
\begin{equation}
\label{eq:Ising_H}
\hat{H}_{\rm Ising} = - h \sum_{j=1}^N \hat{\sigma}_j^x - J \sum_{j=1}^{N-1} \hat{\sigma}_j^z\hat{\sigma}_{j+1}^z \, ,
\end{equation}
where $h>0$ is the intensity of the local fields, $J>0$ is the interaction strength between adjacent spins (sometimes dubbed as exchange constant)~\cite{Sachdev:2011}. We assume, for the sake of simplicity, that $h$ and $J$ are the same across all the $N$ spins of the chain. Note that $\hat{\sigma}_j^k$ (for $k=x,y,z$) are the local Pauli operators acting on the $j$-th spin.

Eq.~\eqref{eq:Ising_H} exhibits a competition between the interaction among adjacent spins in the $z$ direction and the external field in the $x$ direction. 
The external field adds quantum fluctuations to the system, driving the system across a quantum phase transition from an ordered to a disordered phase. 
The TFIM can be analytically solved via Jordan-Wigner transformation \cite{Schulz1964}, showing that -- at zero temperature -- the QPT occurs exactly when $h = J$~\cite{Dziamarga:2005,Sachdev:2011,mbeng:2020}. This critical point separates the ordered ($h<J$) from the disordered ($h>J$) phase, the latter emerging when the transverse field dominates.

Notwithstanding its numerous interesting features, here we consider the quantum Ising model for its metrological properties. In Ref.~\cite{Invernizzi:2008} it has been shown that the accuracy with which we are able to estimate the Hamiltonian parameters $h$ and $J$ is greatly enhanced at critical points with respect to non-critical ones. It is not difficult to show that for a closed quantum Ising chain (at zero temperature) the QFI can be analytically determined via exact diagonalisation of the corresponding Hamiltonian. Interestingly, even for very small sizes (e.g. $N=2,3,4$) the maximum of the QFI is obtained for $h=J$, a result that holds true also in the thermodynamic limit $N\to \infty$~\cite{Invernizzi:2008}.

Differently, here we aim at discussing the possible advantage offered by quantum criticality in estimating the non-Hamiltonian parameter $\lambda$, i.e.~the collapse rate. To this end, we need to let each spin of the chain interact with a local bath mimicking the collapse mechanism, as in Eq.~\eqref{eq:qubit_collapse_dissipator}, whose many-body version reads 
\begin{equation}
\label{eq:Ising_collapse_dissipator}
D_j(\hat{\rho}) = - \frac{\lambda}{2} \, [\hat{\sigma}^z_j, [ \hat{\sigma}^z_j,\hat{\rho}] ] \, ,
\end{equation}
for each site $j=1, \ldots, N$.

Following the non-equilibrium approach put forth in  Sec.~\ref{sec:probing}, we attach a probe (i.e.~a qubit) to the Ising chain, so that in the full Hamiltonian of Eq.~\eqref{eq:non_eq_H_scheme} we take $\hat{H}_S = \hat{H}_{\rm Ising}$, $\hat{H}_{p} = - h_p \hat{\sigma}^x_p$, while the interaction Hamiltonian is given by
\begin{equation}
\label{eq:Ising_probe_int}
\hat{H}_I = - J_p \hat{\sigma}_N^z \otimes \hat{\sigma}_{p}^z \, ,
\end{equation}
where $J_p$ is the coupling constant between the $N$-th spin of the chain and the probe. In other terms, owing to this choice, the composite system described by the Hamiltonian (\ref{eq:non_eq_H_scheme}) can be regarded as a $(N+1)$-dimensional Ising chain, where one site, serving as a probe, is not subject to the collapse mechanism. 

It is important to emphasise that, since $\lambda$ is small compared to the other physical quantities characterising the unitary dynamics, we can work in a low-dissipation regime, where the dissipation induced by Eq.~\eqref{eq:Ising_collapse_dissipator} acts \emph{de facto} as a perturbation~\cite{Rossini:2021}. This allows us to rely on both the critical and metrological properties of the quantum Ising chain, even in the limit of small $N$~\cite{Invernizzi:2008}. Thus, in order to study the estimability of $\lambda$ in relation to the critical behaviour of the Ising model, we work in a situation close to the transition point for the corresponding infinite-dimensional closed system, i.e. $h = J$.

\section{Analysis and Results}
\label{sec:results}

Given the many-body probing setup described in Sec.~\ref{sec:many-body}, we discuss in this section the problem of estimating the collapse rate $\lambda$.

The first step is to solve the master equation for an open many-body system, which can be -- in general -- a formidable task~\cite{Weimer:2021}. In our study we follow a simple approach consisting in vectorising [cf.~Appendix \ref{app:vectorisation}] the Lindblad master equation
\begin{equation}
\label{eq:Ising_collapse_ME}
\frac{\D \hat{\rho}}{\D t} = - i [\hat{H}, \hat{\rho}] + \sum_{j=1}^N D_j(\hat{\rho}) \, ,
\end{equation}
and numerically solving the corresponding set of ordinary differential equations. We should mention that this method scales exponentially with the system size $N$, as the Hilbert space scales as $2^{N+1}$. However, it allows us to gain insight on the issue that we want to investigate, without resorting to more sophisticated simulation techniques.

Specifically, we choose the following initial state $\hat{\rho}(0) = \hat{\rho}_{\rm Ising}(0) \otimes \hat{\rho}_p(0)$, where the Ising chain is in a thermal Gibbs state $\hat{\rho}_{\rm Ising}(0) \equiv \exp(-\beta \hat{H}_{\rm Ising})/\mathcal{Z}_{\rm Ising}$, with $\mathcal{Z}_{\rm Ising} = \tr{[ \exp(-\beta \hat{H}_{\rm Ising}) ]}$, and  $\beta = 1/T$ being the inverse temperature. As in Eq.~\eqref{eq:qubit_initial_state}, the probe is prepared in a pure state in the form $\hat\rho_p = \ket{\psi_p} \! \bra{\psi_p}$. 

As for the single-body case, since we need to compute the QFI at a given time, we have to carefully choose the timescale so that, depending on the value of $\lambda$, we can detect the effect of the collapse mechanism. On one hand, as $\lambda$ gets smaller, it takes longer to see the probe's dynamics departing from the fully coherent oscillations as a result of the collapse. Moreover, the size $N$ of the chain plays a crucial role on the dynamics of the probe.  By increasing $N$, the probe converges quicker to its steady state. A rough estimate of the typical relaxation time is given by $\tau_R \approx 1/\Delta_\mathcal{L}$, where $\Delta_\mathcal{L}$ is the Liouvillian gap  [cf. Appendix \ref{app:vectorisation}].

By tracing out the degrees of freedom associated with the Ising chain only, we obtain the state of the probe at any time $t$, i.e. $\hat{\rho}_p(t) = \partr_S{\hat{\rho}(t)}$. The latter is 
needed to tackle our estimation problem, using it to calculate the QFI by resorting to Eq.~\eqref{eq:qubit_QFI_eig_split}.

To study the behaviour of the QFI as a function of $\lambda$, we proceed as follows. For a given set of physical parameters (including the value of the collapse rate $\lambda$), we calculate the QFI as a function of time, i.e. $G=G(t)$, determining the optimal time $t_{\rm opt}$ such that $G_{\rm max} = G(t_{\rm opt})$. We then fix $t=t_{\rm opt}$ and we study the QFI as a function of $\lambda$, i.e. $G=G(\lambda)$ over an interval of interest $\mathcal{I}_{\lambda} = [\lambda_{\rm min}, \lambda_{\rm max}]$.

To study the behaviour of the QFI as a function of the collapse parameter $\lambda$, i.e. $G=G(\lambda)$, we consider four different intervals for $\lambda$, namely $\mathcal{I}_{\lambda_1} = [0.1,0.5]$, $\mathcal{I}_{\lambda_2} = [10^{-3},10^{-1}]$, $\mathcal{I}_{\lambda_3} = [10^{-5},10^{-3}]$ and $\mathcal{I}_{\lambda_4} = [10^{-7},10^{-5}]$. In each of these intervals we determine $t = t_{\rm opt}$ corresponding to the lower bound of each interval, i.e. $\lambda = \lambda_{\rm min}$.

In our numerical simulations, we test the behaviour of the QFI $G=G(\lambda)$ in three different scenarios: at the critical point (i.e. $h=J$), slightly below or above it (i.e. $h = 0.99 J$ and $h = 1.01 J$, respectively). Furthermore, we study how the QFI scales with the system size by taking $N=2,3,4$ and $5$. 

\begin{figure*}[t!]
\centering
     \begin{minipage}[b]{\columnwidth}
        \centering
         \includegraphics[width=\columnwidth]{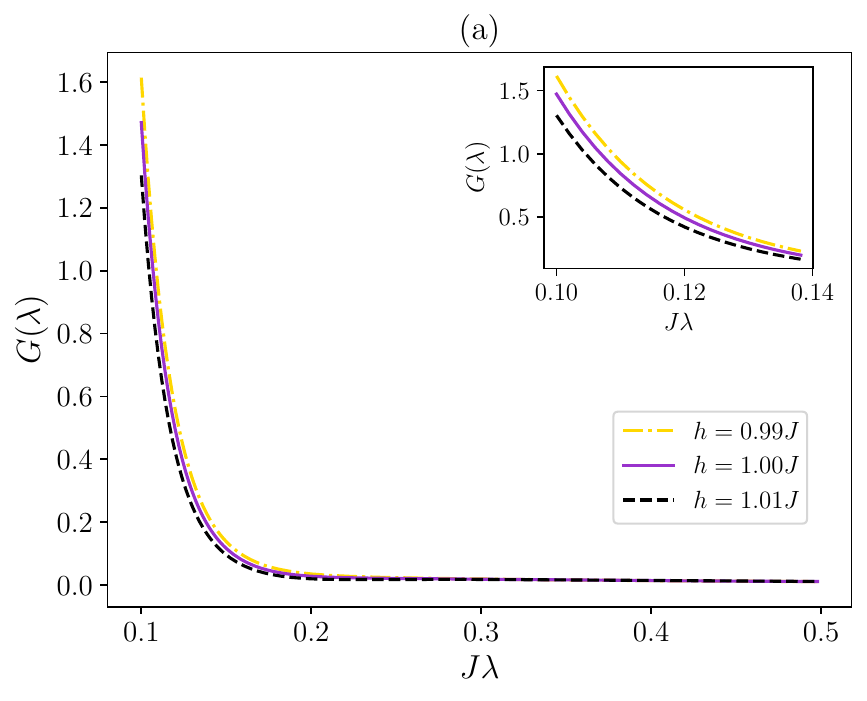}
    \end{minipage}
    \hfill
    \begin{minipage}[b]{\columnwidth}
        \centering
        \includegraphics[width=\columnwidth]{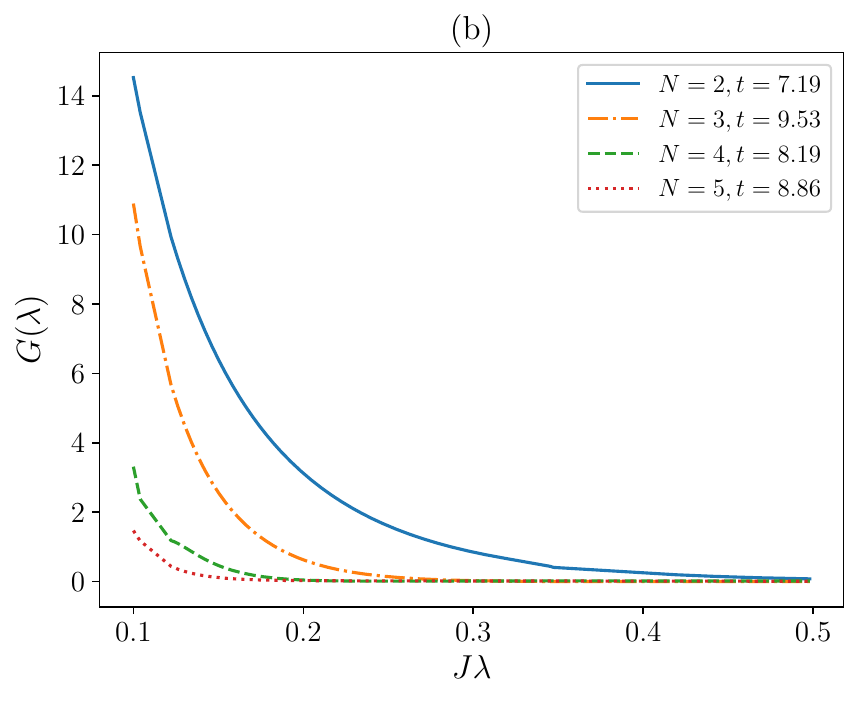}
    \end{minipage}
        \caption{Non-equilibrium probing of a quantum Ising chain: plots of the QFI $G$ of the probe as a function of the collapse rate $\lambda$, when $\lambda \in \mathcal{I}_{\lambda_1} = [0.1,0.5]$. In Panel (a) we consider the case $N=5$. We plot $G$ for three different values of $h$: at the critical point (i.e. $h=J$, solid line), slightly below (i.e. $h=0.99J$, dash-dotted line), and slightly above (i.e. $h=1.01J$, dashed line). In Panel (b), we plot $G$ for different values of the system size $N$ when $h=J$. For the numerical simulations, we chose $h_p = 0.5 J$, $J_p= 0.5J$, while for the initial state, we took $\beta = 0.1/J$, $\theta=\pi$, $\phi=0$. The times at which the QFI is measured, i.e. $t=t_{\rm opt}$ (in units of $J$), are displayed in the legend of Panel (b), as reported in the Table~\eqref{table:Ising_topt} of the Appendix~\eqref{app:data}.}
        \label{fig:Ising_collapse_QFI_lambda01}
\end{figure*}

According to our numerical simulations, in the interval $\mathcal{I}_{\lambda_1}$ (where $\lambda$ is comparable to the relevant energies of the system) there is no evidence that quantum criticality is resourceful for estimating the collapse rate $\lambda$, as one can notice in Fig.~\ref{fig:Ising_collapse_QFI_lambda01} (a). Similarly, our ability to discriminate $\lambda$ is not enhanced by the system size: the QFI actually becomes smaller as the system size increases [cf. Fig.~\ref{fig:Ising_collapse_QFI_lambda01} (b)]. This is likely due to the chosen values of $\lambda$, for which the assumption of small collapse-induced perturbations breaks down. In a sense, the regime where $\lambda \in \mathcal{I}_{\lambda_1}$ challenges the validity of the master equation~\eqref{eq:Ising_collapse_ME}, as we can no longer assume the weak coupling between the system and the effective bath.

\begin{figure}
	\centering
        \includegraphics[width=0.9\columnwidth]{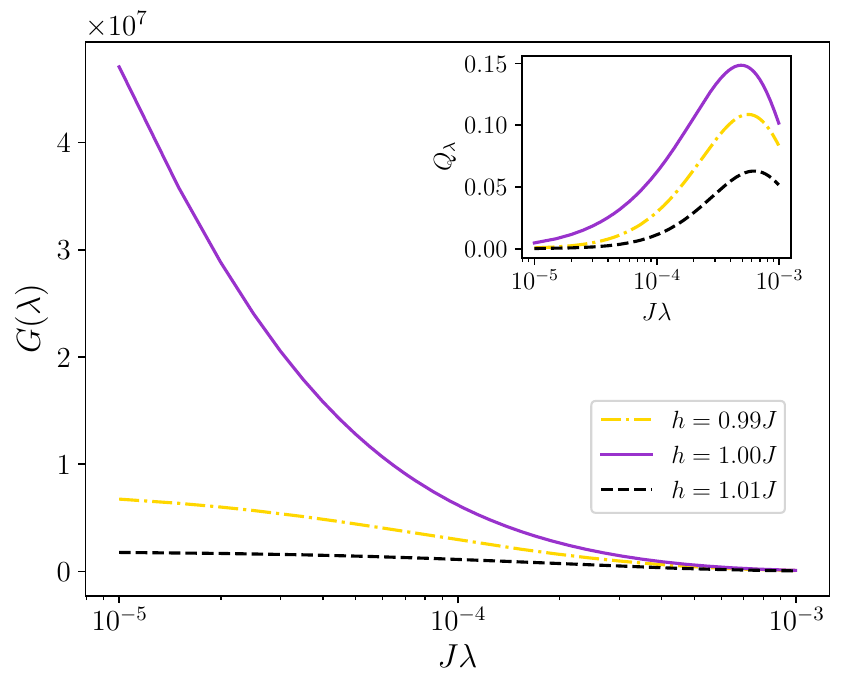}
\caption{QFI as a function of the parameter $\lambda$ when we apply the non-equilibrium probing scheme to an Ising chain, as described in Eq.~\eqref{sec:many-body}. We plot $G$ for three different values of $h$, so that the critical, the ordered vs disordered phases are explored. In these plots, we consider $\lambda \in \mathcal{I}_{\lambda_3} = [10^{-5}, 10^{-3}]$, $N=4$, while all the remaining physical parameters are $h_p = 0.5 J$, $J_p= 0.5J$, $\beta = 0.1/J$, $\theta=\pi$, $\phi=0$. The QFI is evaluated at $t=t_{\rm opt} = 6.84/J \cdot 10^2$. In the inset we show the corresponding QSNR $Q_\lambda$.}
\label{fig:Ising_collapse_criticality}
\end{figure}

At variance, if we consider the parameter regime corresponding to the intervals $\mathcal{I}_{\lambda_j}$ with $j=2,3$ and $4$, we gather numerical evidence that, for a given $N$, quantum criticality helps when we are to estimate $\lambda$. In Fig.~\ref{fig:Ising_collapse_criticality}, we show an explicit example for $\mathcal{I}_{\lambda_3}$. This result is remarkable in that $h/J=1$ is the critical point for infinite system ($N \to \infty$), but nevertheless we see a marked difference between results for $\mid h/J-1 \mid=0.01$ and for $\mid h/J-1\mid=0$ already for the considered small values of $N$.

\begin{figure}
	\centering
        \includegraphics[width=0.9\columnwidth]{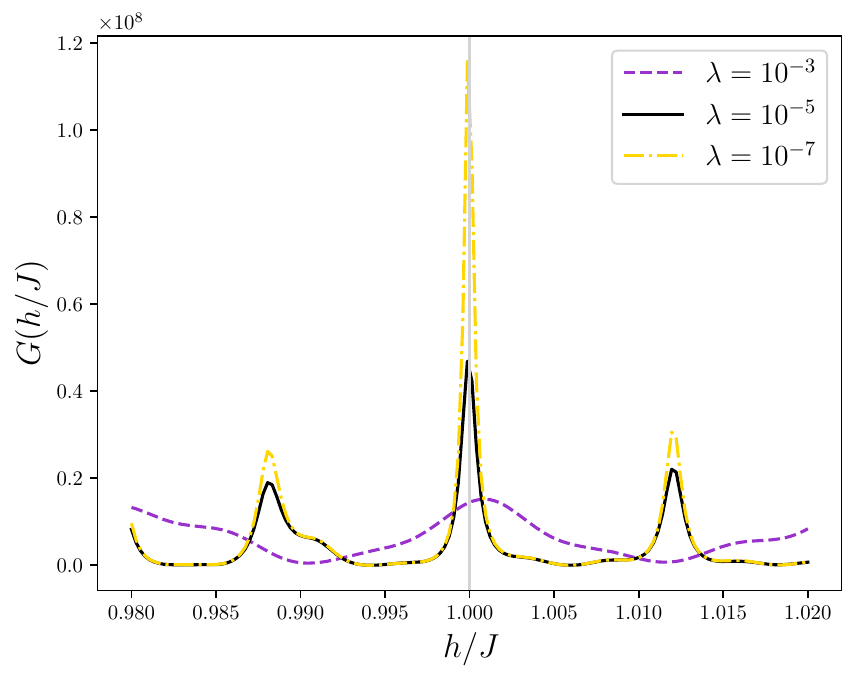}
\caption{QFI as a function of $h/J$ for different values of the collapse rate $\lambda$. In these plots, we consider the case where $N=4$ spins undergo the collapse mechanism. The remaining parameters are $h_p = 0.5 J$, $J_p= 0.5J$, $\beta = 0.1/J$, $\theta=\pi$, $\phi=0$. The QFI is evaluated at $t=t_{\rm opt}$, where the numerical values of $t_{\rm opt}$ are specified in the column corresponding to $N=4$ in the Table~\ref{table:Ising_topt} that can be found in the Appendix~\ref{app:data}. For better visualisation, the curve corresponding to $\lambda = 10^{-3}$ has been rescaled by a factor of $10^2$.}
\label{fig:G_vs_h}
\end{figure}

This goes along the lines of the calculations provided in Ref.~\cite{Invernizzi:2008}, where the authors show that the algebraic expressions for the QFI in a closed quantum Ising chain achieve their maxima at $h=J$ even for $N=2,3,4$. To further elaborate on this point, we study the behaviour of the QFI as a function of the ratio $h/J$, while keeping the collapse rate $\lambda$ fixed. In Fig.~\ref{fig:G_vs_h} we show one instance of this behavior for an Ising chain made of $N=4$ spins. Despite the small size of the chain, the deviations from the theoretical critical point remain relatively small, indicating that operating near the critical condition $h/J=1$ significantly enhances our ability to distinguish $\lambda$. Interestingly, these deviations decrease further for increasingly smaller values of $\lambda$, where dissipative effects are less pronounced, thereby reproducing the results observed in the closed-system case~\cite{Invernizzi:2008}.

\begin{figure*}[t!]
\centering
     \begin{minipage}[b]{0.45\textwidth}
        \centering
         \includegraphics[width=\columnwidth]{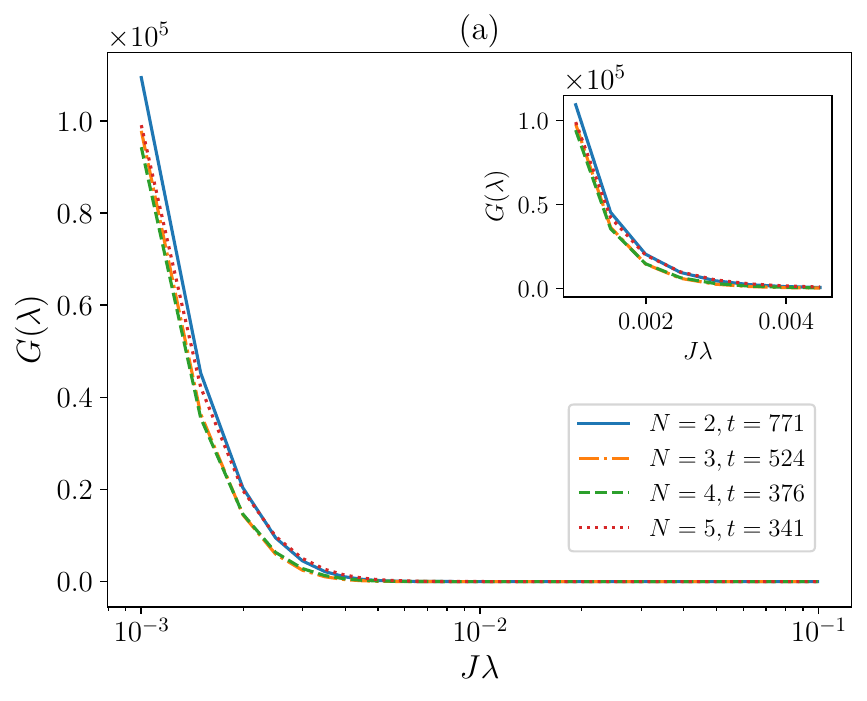}
    \end{minipage}
   \hfill
   \begin{minipage}[b]{0.45\textwidth}
        \centering
         \includegraphics[width=\columnwidth]{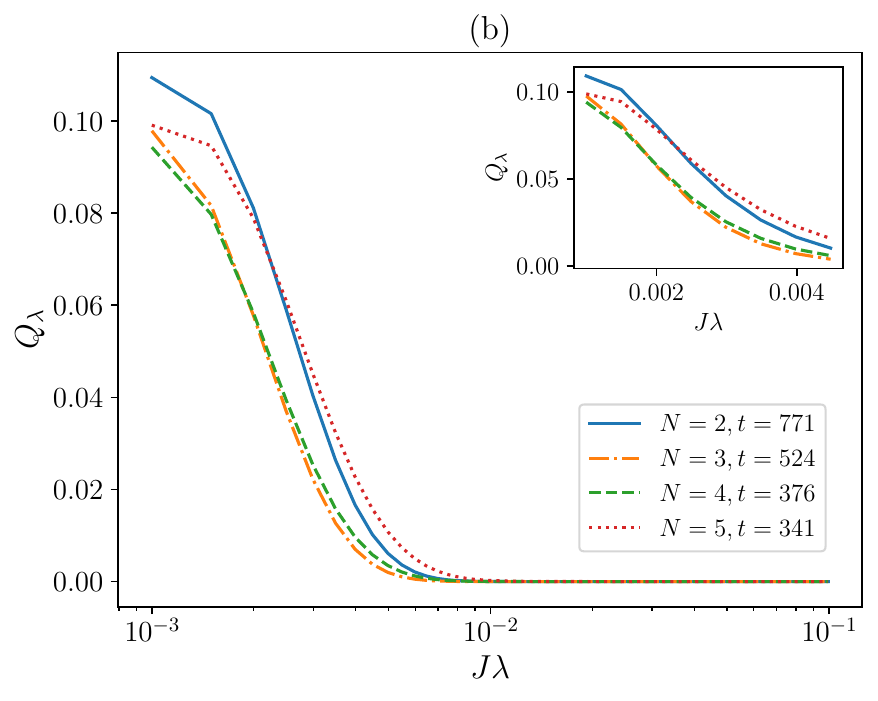}
    \end{minipage}
    \vfill
    \begin{minipage}[b]{0.45\textwidth}
        \centering
        \includegraphics[width=\columnwidth]{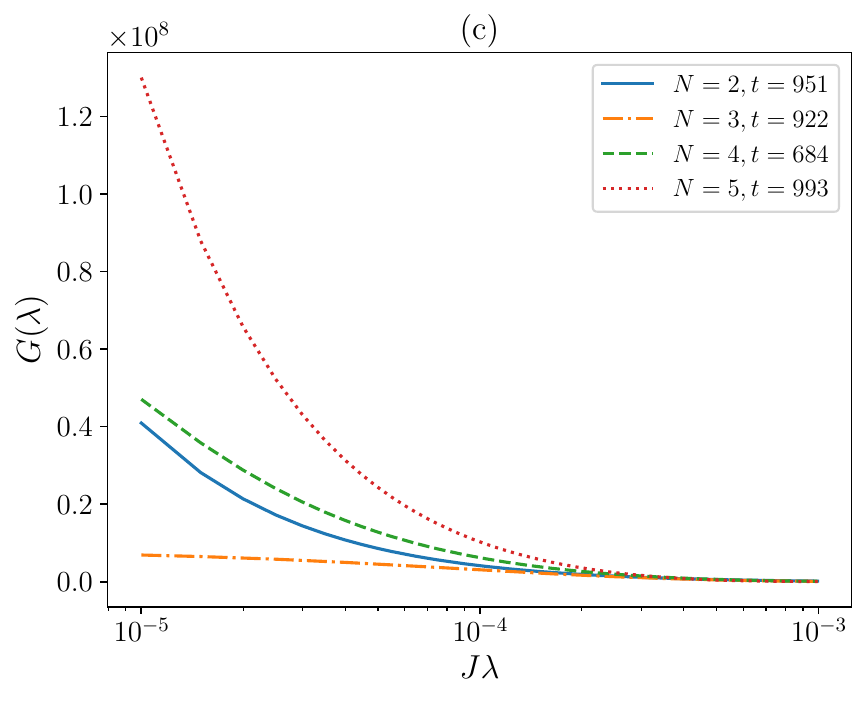}
    \end{minipage}
   \hfill
   \begin{minipage}[b]{0.45\textwidth}
        \centering
         \includegraphics[width=\columnwidth]{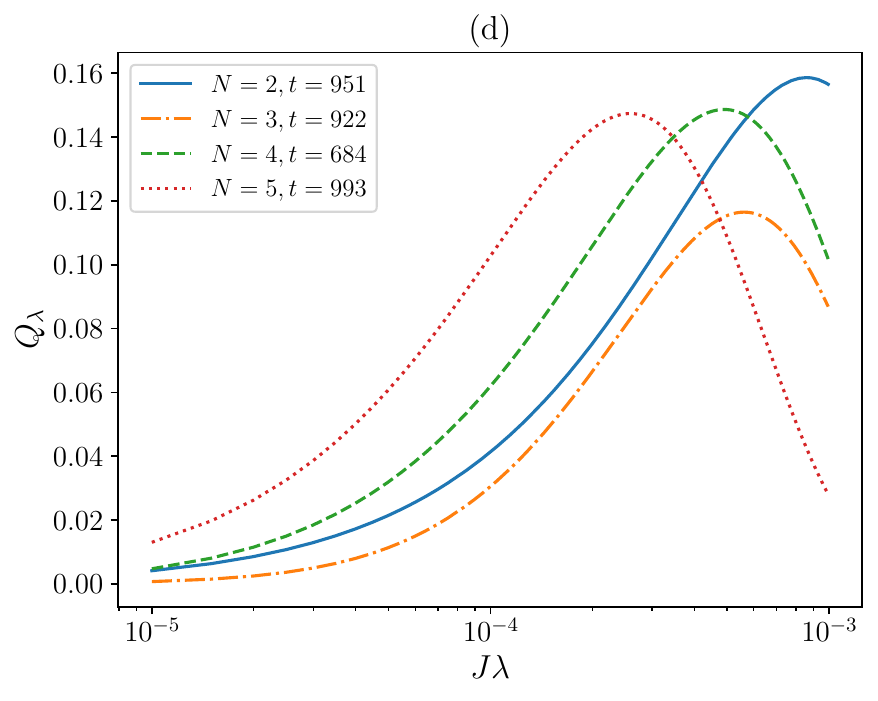}
    \end{minipage}
    \vfill
    \begin{minipage}[b]{0.45\textwidth}
        \centering
        \includegraphics[width=\columnwidth]{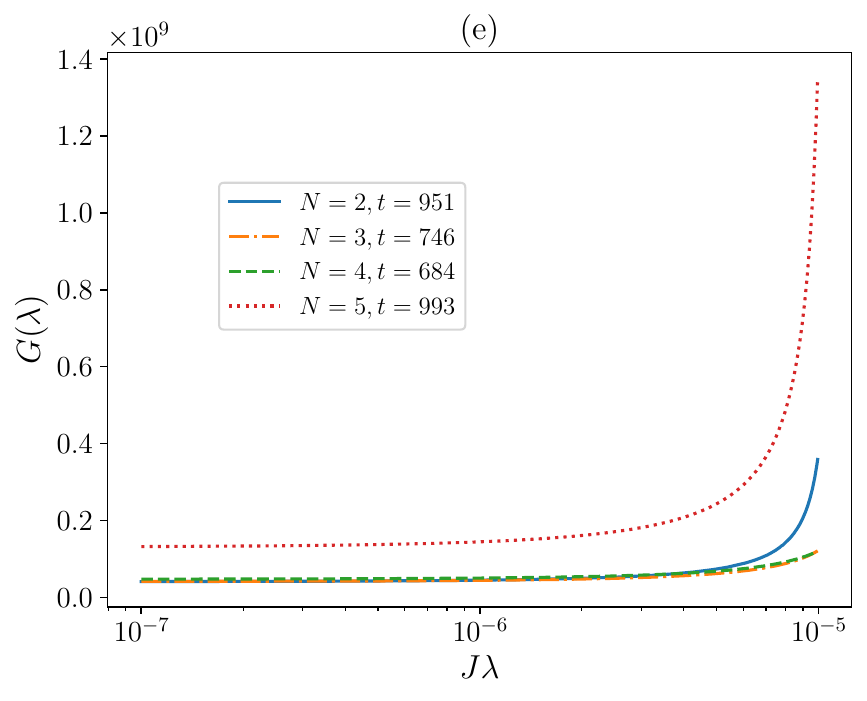}
    \end{minipage}
   \hfill
   \begin{minipage}[b]{0.45\textwidth}
        \centering
         \includegraphics[width=\columnwidth]{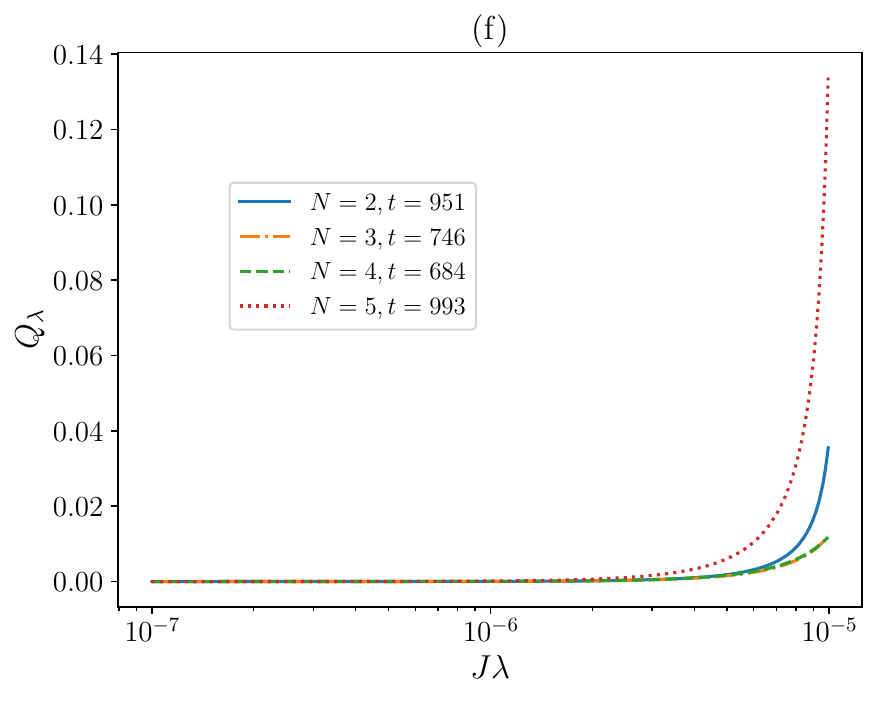}
    \end{minipage}
        \caption{Non-equilibrium probing of a quantum Ising chain: QFI (left column) and QSNR (right column) as a function of $\lambda$, where $\lambda \in \mathcal{I}_{\lambda_j}$, with $j=2$ (first raw), $3$ (second raw), and $4$ (last raw) [Cf. main text] for different system sizes $N$. The numerical simulations are carried out choosing $h_p = 0.5 J$, $J_p= 0.5J$, $\beta = 0.1/J$, $\theta=\pi$, $\phi=0$. The optimal instant of time $t=t_{\rm opt}$ (in units of $J$) for the cases considered above are given in the legend, as well as in Eq.~\eqref{table:Ising_topt} of Eq.~\eqref{app:data}}
        \label{fig:Ising_collpse_QFI_lambda}
\end{figure*}

In the same parameter regimes, our numerical results show that there is no clear monotonicity relationship between the QFI (or the QSNR) and the system size, as one can deduce from the instances reproduced in Fig.~\ref{fig:Ising_collpse_QFI_lambda}. However, if we consider the case corresponding to the interval $\mathcal{I}_{\lambda_3}$ [cf. Fig.~\ref{fig:Ising_collpse_QFI_lambda} (e) and (f)], we can notice that there is a range of values for which the system size plays a positive role in estimating the collapse rate $\lambda$.
Dealing with a larger system appears to enhance the ability to estimate the collapse rate $\lambda$. 
Near the expected critical point (at thermodynamic limit) the correlation length becomes comparable to the system size, corroborating the expectation that, as $N$ grows to thermodynamic limit, the entire system will contribute to amplify the signal. We stress that, although we have used, as probing system, a spin at the edge of the quantum spin chain, the previous argument about the diverging correlation length leads us to conclude that effects qualitatively similar to the ones reported here would hold if the probe is coupled to one spin (or possibly all of the spins) in the bulk of the chain itself. 

\begin{figure}
	\centering
        \includegraphics[width=0.9\columnwidth]{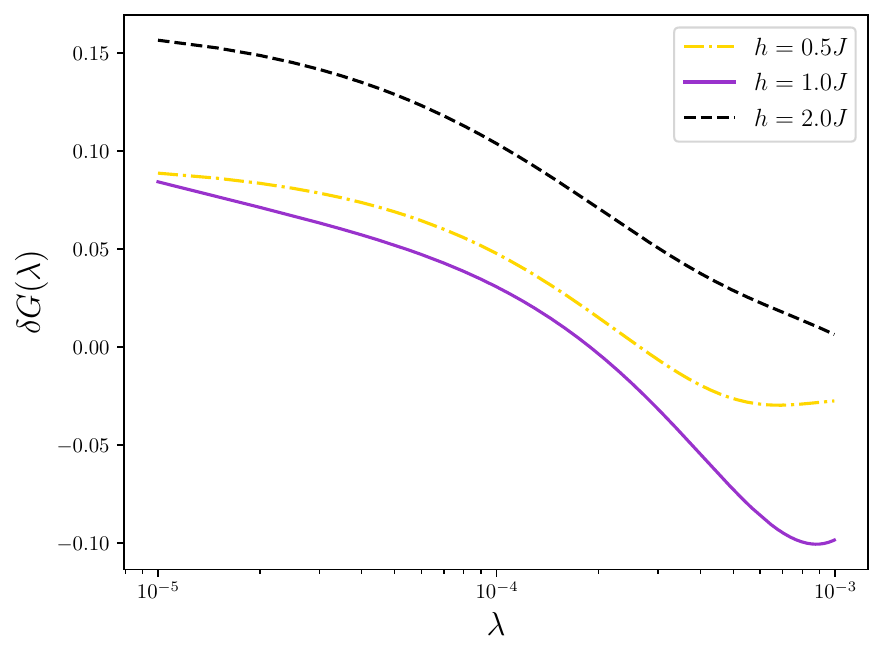}
\caption{Relative difference $\delta G$ of the QFI in presence of correlated noise as described by Eqs.~\eqref{eq:dissipator_corr_noise} and \eqref{eq.f} against the case where the latter is absent so that the dissipator reduces to the one in Eq.~\eqref{eq:Ising_collapse_dissipator}. In the numerical simulations, we consider an Ising chain made of $N=4$ spins, while $\lambda \in \mathcal{I}_{\lambda_3} = [10^{-5}, 10^{-3}]$. All other parameters are $h_p = 0.5 J$, $J_p= 0.5J$, $\beta = 0.1/J$, $\theta=\pi$, $\phi=0$, $t=t_{\rm opt} = 6.84/J \cdot 10^2$, and $r_\text{\tiny C}/a=2$.}
\label{fig:corr_vs_uncorr}
\end{figure}

Furthemore, we can comment on the structure of the collapse dissipator in Eq.~\eqref{eq:Ising_collapse_ME}, which is the incoherent sum of local channels. One can interpret this as if the spatial correlation of the noise mechanisms acting on the elements of the chain has a typical scale $r_\text{\tiny C}$ smaller than the sites interval distance $a$. We can go beyond this assumption by considering a model where such a correlation is introduced. A possibility is to consider  
\begin{equation}
\frac{\D \hat{\rho}}{\D t} = - i [\hat{H}, \hat{\rho}] + \sum_{i,j=1}^N \tilde D_{ij}(\hat{\rho}) \, ,
\end{equation}
where
\begin{equation}
\label{eq:dissipator_corr_noise}
\tilde D_{ij}(\hat{\rho}) = - \frac{\lambda}{2} f(i,j)\, [\hat{\sigma}^z_i, [ \hat{\sigma}^z_j,\hat{\rho}] ] 
\end{equation}
with $f(i,j)$ describing the spatial correlation of the noise between the sites $i$ and $j$.
An example of this correlation function can be borrowed from the CSL model \cite{carlesso2022present}, i.e.
\begin{equation}    \label{eq.f}
    f(i,j)=e^{-a^2(i-j)^2/4r_\text{\tiny C}^2},
\end{equation}
which correlates sites $i$ and $j$ if their relative distance $a(i-j)$ is smaller than the correlation length $r_\text{\tiny C}$. Conversely, if the distance is larger than $r_\text{\tiny C}$, the noise acts incoherently as described in Eq.~\eqref{eq:Ising_collapse_dissipator}. To quantify the effect of the correlated noise, we can explicitly compute the relative difference  $\delta G = (G_{\text{corr}}- G_{\text{uncorr}})/G_{\text{corr}}$, where $G_{\text{uncorr}}$ and $G_{\text{corr}}$ is the QFI computed when the dissipator entering in the Lindblad master equation is given by Eqs.~\eqref{eq:Ising_collapse_dissipator} and~\eqref{eq:dissipator_corr_noise}, respectively. In general, no significant changes are observed when one considers the case of correlated noise, as emerges, for instance, in the case plotted in Fig.~\ref{fig:corr_vs_uncorr}.

\section{Conclusions}
\label{sec:conclusions}

We have introduced a quantum probing approach to the estimation of the collapse rate affecting a quantum critical system. We have considered the case of a quantum TFIM where each site is affected by the collapse mechanism. We first considered the case where the latter mechanism acts independently on each site of the chain. We then considered the case of correlated noise between pairs of sites of the chain, showing that, despite affecting the system dynamics, such a dissipator does not qualitatively change the overall picture in our estimation problem. 
We have provided strong numerical evidence that, in the model under study, the long-range correlations established in proximity of the expected critical point can result in an enhancement of the sensitivity of the QFI of a quantum probe, thus providing an enhanced measurement precision. 

On a general note, we have introduced a non-equilibrium probing scheme for estimating a non-Hamiltonian parameter entering in the dynamical equations of a dissipative Ising chain. More specifically, we provide an accessible way to probe the collapse mechanism in a quantum Ising chain, where criticality plays a crucial role. This version of the collapse mechanism, yet simplified, helps to gain insight on the issue, which can be further improved by considering other many-body systems, as well as more sophisticated versions of the collapse mechanism, such as the mass proportional CSL model~\cite{Bassi_rev:2003}.

Furthermore, we restricted our attention to the case of one single parameter to estimate, i.e., $\lambda$. In principle, one can generalise the problem through a multiparameter approach~\cite{Demkowicz-Dobrzanski:2020,Albarelli:2020}, where more than one parameter need to be estimated, e.g., those characterising the initial state of the probe or the interaction with it. Though this would make the problem  mathematically more demading, it would narrow the gap between our theoretical proposal and its possible experimental implementations. The latter would also require an accurate analysis about the way to drive the system towards the critical point, including potential issues due to the so-called critical slowing down.

Our work paves the way to the further exploration the exclusion plot for proposed collapse models through criticality-assisted quantum metrology, and thus provides a new set of tools in the quest for the falsification of quantum collapse models.

\acknowledgements
GZ thanks the University of Trieste for hospitality during the early stages of this project. We acknowledge support by the European Union's Horizon Europe EIC-Pathfinder project QuCoM (101046973), 
the Leverhulme Trust Research Project Grant UltraQuTe (grant RPG-2018-266), the Royal Society Wolfson Fellowship (RSWF/R3/183013), the UK EPSRC (EP/T028424/1), the Department for the Economy Northern Ireland under the US-Ireland R\&D Partnership Programme, the Italy-UK CNR/RS Joint Project ``Testing fundamental theories with ultracold atoms'', and the PNRR PE National Quantum Science and Technology Institute (PE0000023).

\appendix

\section{Vectorization method}
\label{app:vectorisation}

In this Appendix, we summarise the so-called \emph{vectorization} operation that maps a matrix $\mathbf{M}$ onto a vector $\ve{\mathbf{M}}=\vec{M}$ in the following way:
\begin{equation}
\label{eq:vec1}
\operatorname{vec} \begin{pmatrix} a & b \\ c & d \end{pmatrix} = \begin{pmatrix} a \\ c \\ b \\ d  \end{pmatrix} \, .
\end{equation}
It is useful to introduce the following property~\cite{Turkington:2013}:
\begin{equation}
\label{eq:vec_prop}
\ve{\mathbf{A}\mathbf{B}\mathbf{C}} = \left ( \mathbf{C}^T \otimes \mathbf{A} \right ) \ve{\mathbf{B}} \, ,
\end{equation}
given three matrices $\mathbf{A},\mathbf{B},\mathbf{C}$.

This change of basis is particularly suitable  for solving the dynamics of an open quantum system, when it is governed by a master equation of the form~\cite{Breuer-Petruccione,Manzano:2020}
\begin{equation}
\label{eq:ME_GKSL}
\frac{\D \hat{\rho}}{\D t} = - i  [ \hat{H} , \hat{\rho} ] + D(\hat{\rho}) \, , 
\end{equation}
where the dissipator is in the GKSL form~\cite{Lindblad:1976,Gorini:1976,Gorini:1978}
\begin{equation}
\label{eq:GKSL_diss}
D(\hat{\rho}) = \sum_{k} \gamma_k \left ( \hat{L}_k \hat{\rho} \, \hat{L}_k^\dagger - \frac{1}{2} \left \{ \hat{L}_k^\dagger \hat{L}_k, \hat{\rho}\right \}\right ) \, ,
\end{equation}
with $\gamma_k > 0$ being the relaxation rates, and $\{ \hat{L}_k \}_k$ the set of Lindblad operators.

The main idea behind vectorisation is to bring Eq.~\eqref{eq:ME_GKSL} into a first-order ordinary differential equation in the matrix form $\D\mathbf{x}/\D t = \mathbf{A} \mathbf{x}$. Note that, if the system is a single qubit, the vectorisation of the density matrix consist in mapping
\begin{equation}
\label{eq:vec_unitary}
\hat{\rho} = \begin{pmatrix} \rho_{00} & \rho_{01} \\
\rho_{10} & \rho_{11} \end{pmatrix}
\longmapsto \lvec{\rho} = \begin{pmatrix} \rho_{00} \\
\rho_{10} \\ \rho_{01} \\ \rho_{11}
\end{pmatrix} \, ,
\end{equation}
which is a direct application of Eq.~\eqref{eq:vec1}. 
In order to vectorise Eq.~\eqref{eq:ME_GKSL}, we can resort to the property (\ref{eq:vec_prop}). The right-hand side is readily vectorised by inserting the identity operator $\mathbbm{1}$ where needed. For the unitary part, we have
\begin{align}
\label{eq:vec_non-unitary}
- i & \left ( \hat{H} \hat{\rho} \mathbbm{1} - \mathbbm{1} \hat{\rho} \hat{H} \right )   \mapsto \vec{\mathcal{L}}_U \lvec{\rho} \equiv - i \left ( \mathbbm{1} \otimes \hat{H} - \hat{H}^T \otimes \mathbbm{1} \right ) \lvec{\rho} \, .
\end{align}
Similarly, the dissipator can be vectorised as
\begin{align}
&\sum_{k}  {\gamma_k} \left ( \hat{L}_k \hat{\rho} \, \hat{L}_k^\dagger - \frac{1}{2} \left ( \hat{L}_k^\dagger \hat{L}_k \hat{\rho} \mathbbm{1} + \mathbbm{1} \hat{\rho} \hat{L}_k^\dagger \hat{L}_k \right )\right ) \mapsto \\
\nonumber
&\vec{\mathcal{L}}_D \lvec{\rho} {\equiv} \sum_{k} \gamma_k \left[ \hat{L}_k^* {\otimes} \hat{L}_k \nonumber {-} \frac{1}{2} \left ( \mathbbm{1} \otimes \hat{L}_k^\dagger \hat{L}_k {+} (\hat{L}_k^\dagger \hat{L}_k )^T {\otimes} \mathbbm{1} \right ) \right ] \lvec{\rho} \, .
\end{align}
Therefore, Eq.~\eqref{eq:ME_GKSL} becomes 
\begin{equation}
\label{eq:ME_GKSL_vec}
\frac{\D \lvec{\rho_t}}{\D t} = \vec{\mathcal{L}} \lvec{\rho_t} \, ,
\end{equation}
where $\vec{\mathcal{L}} \equiv \vec{\mathcal{L}}_U + \vec{\mathcal{L}}_D$. The formal solution of Eq.~\eqref{eq:ME_GKSL_vec} is obtained by exponentiation of the time-independent Liouvillian $\vec{\mathcal{L}}$, i.e. 
\begin{equation}
\label{eq:ME_GKSL_vec_sol}
\lvec{\rho_t} = e^{t \vec{\mathcal{L}}} \lvec{\rho_0} \, .
\end{equation}
Therefore, solving the dynamics of an open quantum system described by the Eq.~\eqref{eq:ME_GKSL} is equivalent to exponentiating a complex, non-Hermitian matrix $\vec{\mathcal{L}}$~\cite{Landi_rev:2022,Campaioli:2024}. The vectorised Liouvillian $\vec{\mathcal{L}}$ possesses two distinct families of right and left eigenvectors, i.e. $\{\lvec{R_\alpha}\}_{\alpha}$ and $\{\lvec{L_\alpha}\}_\alpha$, respectively, associated with the singular value $l_\alpha$, sometimes dubbed as \emph{rapidity}, the latter being in general a complex number. The right and left eigenvectors are defined by the equations 
$\vec{\mathcal{L}} \lvec{R_\alpha}  = l_\alpha \lvec{R_\alpha}$ and 
$\langle \langle {L}_{\alpha} | \vec{\mathcal{L}} = l_\alpha \langle \langle {L}_{\alpha} | $, respectively.

When vectorising a master equation, we are actually working in a Hilbert space endowed with the scalar product $\langle \langle \phi | \chi \rangle \rangle \equiv \tr{[\hat{\phi}^\dagger \hat{\chi} ]}$. Note that the right and left eigenvectors are not normalised; they can easily normalised by setting $| \tilde{R}_\alpha \rangle \rangle = \lvec{R_\alpha}/\sqrt{\langle \langle L_\alpha | R_\alpha \rangle \rangle}$ and $\langle \langle \tilde{L}_\alpha | = \langle \langle L_\alpha|/\sqrt{\langle \langle L_\alpha | R_\alpha \rangle \rangle}$, so that they automatically fulfil the orthonormality condition $\langle \langle \tilde{L}_\alpha | \tilde{R}_{\alpha'} \rangle \rangle = \delta_{\alpha \alpha'}$. 
We can decompose the full Liouvillian as
\begin{equation}
\label{eq:Liouville_spectral_dec}
\vec{\mathcal{L}} = \sum_\alpha l_\alpha | \tilde{R}_\alpha \rangle \rangle \langle \langle \tilde{L}_\alpha | \, .
\end{equation}
Starting from Eq.~\eqref{eq:Liouville_spectral_dec}, one can expand the exponential and resort to the condition $\langle \langle \tilde{L}_\alpha | \tilde{R}_{\alpha'} \rangle \rangle = \delta_{\alpha \alpha'}$ to obtain 
\begin{equation}
    \label{eq:Ising_ev_exp}
    e^{\hat{\mathcal{L}}t} = \sum_\alpha e^{l_\alpha t} | \tilde{R}_\alpha \rangle \rangle \langle \langle \tilde{L}_\alpha | \, .
\end{equation}
Combining Eq.~\eqref{eq:ME_GKSL_vec_sol} and \eqref{eq:Ising_ev_exp}, we get
\begin{equation}
    \label{eq:Ising_Liouville_sol_eq}
    \lvec{\rho_t} = \sum_\alpha c_\alpha e^{l_\alpha t} | \tilde{R}_\alpha \rangle \rangle \, ,
\end{equation}
where $c_\alpha \equiv \langle \langle \tilde{L}_\alpha | \rho_0\rangle \rangle $ are determined by the initial conditions. Essentially, the evolved state $\lvec{\rho_t}$ is given a linear combination of the right eigenvectors $| \tilde{R}_\alpha \rangle \rangle$ with certain coefficients $c_\alpha$, while the time dependence is all contained in the exponentials $e^{l_\alpha t}$.

The spectrum of the Liouvillians contains interesting information about the system's dynamics~\cite{Albert:2014,Albert:2016}. Specifically, we can summarise some of the properties of $\vec{\mathcal{L}}$ as follows~\cite{Chruscinski:2022}
\begin{enumerate}[label=(\roman*)]
  \item the spectrum $\{l_\alpha\}_{\alpha}$ is symmetric with respect to the real line,
  \item all eigenvalues $l_\alpha$ belong to the left half of the complex plane, i.e. $\mathbb{C}_{-} = \{ z \in \mathbb{C} | \operatorname{Re} z \le 0 \}$,
  \item the right eigenvector corresponding to the eigenvalue $l_0 = 0$, namely $|\tilde{R}_0 \rangle \rangle$, defines the steady state, i.e. $\vec{\mathcal{L}} \lvec{{\rho}_{\rm ss}} = \vec{\mathcal{L}} |\tilde{R}_0 \rangle \rangle = 0$.
\end{enumerate}

One more property of the Liouvillian spectrum allows us to estimate how fast a quantum system approaches the stationary state when is dissipatively coupled to an external environment. To this end, it is worth mentioning that the rapidities, being in general complex, can be denoted as $l_\alpha = - i \omega_\alpha - \Gamma_\alpha$,
with $\omega_\alpha \equiv -\operatorname{Im} l_\alpha$ and $\Gamma_\alpha \equiv - \operatorname{Re} l_\alpha$.
The set $\{\Gamma_\alpha\}_\alpha$ defines the relaxation rates. This is justified by close inspection of Eq.~\eqref{eq:Ising_Liouville_sol_eq}, where the rapidities $l_\alpha$ enter: it is immediate to classify the decay mode $\alpha$ into oscillatory [if $\Gamma_\alpha =0$], purely decaying [if $\omega_\alpha =0$], decaying spirals [if $\Gamma_\alpha > 0$ and $\omega_\alpha\ne 0$]~\cite{Albert:2014}.
In particular, we can define the so-called \emph{Liouvillian gap} as $\Delta_\mathcal{L} \equiv \min_{\alpha\ne 0} \Gamma_\alpha$, which identifies the decay rate of the slowest relaxation mode~\cite{Spohn:1976}. It is reasonable to give $1/\Delta_\mathcal{L}$ the meaning of an upper bound to the relaxation time. While such interpretation will hold when considering many-body systems~\cite{Rossini:2021,Carollo:2020}, 
it is worth mentioning that there are some instances contradicting it~\cite{Mori:2020,Mori:2023}.

\section{Vectorisation of Eq.~\eqref{eq:1qubit_ME}}
\label{app:sol_ME}

The results summarised in Appendix~\ref{app:vectorisation} can be applied to vectorise Eq.~\eqref{eq:1qubit_ME}. We have that Eq.~\eqref{eq:vec_unitary} and \eqref{eq:vec_non-unitary} yield 
\begin{align}
\vec{\mathcal{L}}_U & = - i \left ( \mathbbm{1}_2 \otimes \hat{H} - \hat{H} \otimes \mathbbm{1}_2  \right ) \, , \\ 
\vec{\mathcal{L}}_D & = \lambda \left ( \hat{\sigma}^z \otimes \hat{\sigma}^z - \mathbbm{1}_4\right) \, ,
\end{align}
where $\mathbbm{1}_2$ and $\mathbbm{1}_4$ are the identity operators defined over the two- and four-dimensional Hilbert space, respectively. The full Liouvillian $\vec{\mathcal{L}} = \vec{\mathcal{L}}_U + \vec{\mathcal{L}}_D$ can be cast as the non-Hermitian matrix 
\begin{equation}
\vec{\mathcal{L}} = \frac{1}{2}
\begin{pmatrix}
0 & - i \omega_0 & i \omega_0 &  0 \\
- i \omega_0 & - 4 \lambda & 0 & i \omega_0 \\
i \omega_0 & 0 & - 4 \lambda &  - i \omega_0 \\
0 & i \omega_0 & - i \omega_0 & 0
\end{pmatrix} \, ,
\end{equation}
which can be diagonalised to get the following eigenvalues
\begin{equation}
\label{eq:collapse_eigenvalues}
l_0 = 0, \quad l_1= -2 \lambda, \quad l_\pm = \ \lambda \pm \sqrt{\lambda^2 - \omega_0^2} \, ,
\end{equation}
while the corresponding (unnormalised) right eigenvectors read as
\begin{equation}\begin{aligned}
\label{eq:collapse_eigenvectors}
\lvec{R_0} = & \begin{pmatrix} 1 \\ 0 \\ 0  \\ 1 \end{pmatrix} \, , \quad
\lvec{R_1} = \begin{pmatrix} 0 \\ 1 \\ 1  \\ 0 \end{pmatrix} \, ,  \\
\lvec{R_\pm} = \frac{1}{\omega_0}& \begin{pmatrix} -\omega_0 \\  {i} \left ( \lambda \mp \sqrt{ - \lambda^2 + \omega_0^2}\right) \\ i \left ( -\lambda \pm \sqrt{ - \lambda^2 + \omega_0^2}\right)  \\ \omega_0 \end{pmatrix}.
\end{aligned}\end{equation}
 If $\omega_0 \ne \lambda$, all the eigenvalues are simple and the corresponding quantum dynamical semigroup $\{ e^{t \mathcal{L}} \}_{t \ge 0}$ is relaxing towards a unique maximally mixed state~\cite{Evans:1977,Nigro:2019,Chruscinski:2022,Zhang:2024}. The steady state $\hat\rho_{\rm ss}$ can be readily found considering the eigenvector corresponding to the eigenvalue $l_0=0$, i.e. $\lvec{R_0}$. Upon de-vectorisation and normalisation, we obtain the totally mixed state
\begin{align}
\hat{\rho}_{\rm ss} = \frac{1}{2} \mathbbm{1}_2.
\end{align}
In general, the state at a time $t$ is given by
\begin{equation}
\lvec{\rho_t} = \sum_{\alpha=0}^3 c_{\alpha} e^{l_\alpha t} \lvec{R_\alpha},~~\text{with}~~c_\alpha = \langle \langle L_\alpha | \rho_0 \rangle \rangle,
\end{equation}
together with Eqs.~\eqref{eq:collapse_eigenvalues} and \eqref{eq:collapse_eigenvectors}. 
    
\section{Further data about numerical simulations}
\label{app:data}

In this Appendix, we show some numerical data regarding the quantum Ising chain discussed in Secs.~\ref{sec:many-body} and \ref{sec:results}.
For studying the behaviour of the QFI against the collapse rate $\lambda$, we consider the regime in which $h_p = 0.5 J$, $J_p= 0.5J$, $\beta = 0.1/J$, $\theta=\pi$, $\phi=0$. In Table ~\ref{table:Ising_topt} we report the optimal time $t=t_{\rm opt}$ at which the QFI is determined in the examples provided in Sec.~\ref{sec:results}. 

\renewcommand{\arraystretch}{1.3}
\begin{table}[h!]
\begin{center}
\begin{tabular}{ c|c|c|c|c } 
\diagbox{$J\lambda$}{$N$}  & $2$  &  $3$ &  $4$ &  $5$ \\  [0.5ex]
 \hline \hline
$10^{-1}$ & $7.19$ & $9.53$ & $8 .19$ & $8.86$ \\  [0.5ex]
  \hline
$10^{-3}$ & $7.71 \times 10^{2}$ & $5.24 \times 10^{2}$ & $3.76 \times 10^{2}$ & $2.41 \times 10^{2}$ \\  [0.5ex]
\hline
$10^{-5}$ & $9.51 \times 10^{2}$ & $9.22 \times 10^{2}$ & $6.84 \times 10^{2}$ & $9.93 \times 10^{2}$ \\ 
\hline
$10^{-7}$ & $9.51 \times 10^{2}$ & $7.46 \times 10^{2}$ & $6.84 \times 10^{2}$ & $9.93 \times 10^{2}$ \\ 
\end{tabular}
\end{center}
\caption{Numerical values of $t_{\rm opt}/J$ for a given value of $\lambda$ and $N$.}
\label{table:Ising_topt}
\end{table}

\bibliography{biblio.bib}

\begin{thebibliography}{78}%
\makeatletter
\providecommand \@ifxundefined [1]{%
 \@ifx{#1\undefined}
}%
\providecommand \@ifnum [1]{%
 \ifnum #1\expandafter \@firstoftwo
 \else \expandafter \@secondoftwo
 \fi
}%
\providecommand \@ifx [1]{%
 \ifx #1\expandafter \@firstoftwo
 \else \expandafter \@secondoftwo
 \fi
}%
\providecommand \natexlab [1]{#1}%
\providecommand \enquote  [1]{``#1''}%
\providecommand \bibnamefont  [1]{#1}%
\providecommand \bibfnamefont [1]{#1}%
\providecommand \citenamefont [1]{#1}%
\providecommand \href@noop [0]{\@secondoftwo}%
\providecommand \href [0]{\begingroup \@sanitize@url \@href}%
\providecommand \@href[1]{\@@startlink{#1}\@@href}%
\providecommand \@@href[1]{\endgroup#1\@@endlink}%
\providecommand \@sanitize@url [0]{\catcode `\\12\catcode `\$12\catcode
  `\&12\catcode `\#12\catcode `\^12\catcode `\_12\catcode `\%12\relax}%
\providecommand \@@startlink[1]{}%
\providecommand \@@endlink[0]{}%
\providecommand \url  [0]{\begingroup\@sanitize@url \@url }%
\providecommand \@url [1]{\endgroup\@href {#1}{\urlprefix }}%
\providecommand \urlprefix  [0]{URL }%
\providecommand \Eprint [0]{\href }%
\providecommand \doibase [0]{http://dx.doi.org/}%
\providecommand \selectlanguage [0]{\@gobble}%
\providecommand \bibinfo  [0]{\@secondoftwo}%
\providecommand \bibfield  [0]{\@secondoftwo}%
\providecommand \translation [1]{[#1]}%
\providecommand \BibitemOpen [0]{}%
\providecommand \bibitemStop [0]{}%
\providecommand \bibitemNoStop [0]{.\EOS\space}%
\providecommand \EOS [0]{\spacefactor3000\relax}%
\providecommand \BibitemShut  [1]{\csname bibitem#1\endcsname}%
\let\auto@bib@innerbib\@empty
\bibitem [{\citenamefont {Bassi}\ and\ \citenamefont
  {Ghirardi}(2003)}]{Bassi_rev:2003}%
  \BibitemOpen
  \bibfield  {author} {\bibinfo {author} {\bibfnamefont {Angelo}\ \bibnamefont
  {Bassi}}\ and\ \bibinfo {author} {\bibfnamefont {GianCarlo}\ \bibnamefont
  {Ghirardi}},\ }\bibfield  {title} {\enquote {\bibinfo {title} {Dynamical
  reduction models},}\ }\href {\doibase
  https://doi.org/10.1016/S0370-1573(03)00103-0} {\bibfield  {journal}
  {\bibinfo  {journal} {Physics Reports}\ }\textbf {\bibinfo {volume} {379}},\
  \bibinfo {pages} {257--426} (\bibinfo {year} {2003})}\BibitemShut {NoStop}%
\bibitem [{\citenamefont {Ghirardi}\ \emph {et~al.}(1990)\citenamefont
  {Ghirardi}, \citenamefont {Pearle},\ and\ \citenamefont
  {Rimini}}]{ghirardi1990markov}%
  \BibitemOpen
  \bibfield  {author} {\bibinfo {author} {\bibfnamefont {Gian~Carlo}\
  \bibnamefont {Ghirardi}}, \bibinfo {author} {\bibfnamefont {Philip}\
  \bibnamefont {Pearle}}, \ and\ \bibinfo {author} {\bibfnamefont {Alberto}\
  \bibnamefont {Rimini}},\ }\bibfield  {title} {\enquote {\bibinfo {title}
  {Markov processes in hilbert space and continuous spontaneous localization of
  systems of identical particles},}\ }\href
  {https://doi.org/10.1103/PhysRevA.42.78} {\bibfield  {journal} {\bibinfo
  {journal} {Physical Review A}\ }\textbf {\bibinfo {volume} {42}},\ \bibinfo
  {pages} {78} (\bibinfo {year} {1990})}\BibitemShut {NoStop}%
\bibitem [{\citenamefont {Di{\'o}si}(1989)}]{diosi1989models}%
  \BibitemOpen
  \bibfield  {author} {\bibinfo {author} {\bibfnamefont {Lajos}\ \bibnamefont
  {Di{\'o}si}},\ }\bibfield  {title} {\enquote {\bibinfo {title} {Models for
  universal reduction of macroscopic quantum fluctuations},}\ }\href
  {https://doi.org/10.1103/PhysRevA.40.1165} {\bibfield  {journal} {\bibinfo
  {journal} {Physical Review A}\ }\textbf {\bibinfo {volume} {40}},\ \bibinfo
  {pages} {1165} (\bibinfo {year} {1989})}\BibitemShut {NoStop}%
\bibitem [{\citenamefont {Penrose}(1996)}]{penrose1996gravity}%
  \BibitemOpen
  \bibfield  {author} {\bibinfo {author} {\bibfnamefont {Roger}\ \bibnamefont
  {Penrose}},\ }\bibfield  {title} {\enquote {\bibinfo {title} {On gravity's
  role in quantum state reduction},}\ }\href
  {https://doi.org/10.1007/BF02105068} {\bibfield  {journal} {\bibinfo
  {journal} {General relativity and gravitation}\ }\textbf {\bibinfo {volume}
  {28}},\ \bibinfo {pages} {581--600} (\bibinfo {year} {1996})}\BibitemShut
  {NoStop}%
\bibitem [{\citenamefont {Bassi}\ \emph {et~al.}(2013)\citenamefont {Bassi},
  \citenamefont {Lochan}, \citenamefont {Satin}, \citenamefont {Singh},\ and\
  \citenamefont {Ulbricht}}]{bassi2013models}%
  \BibitemOpen
  \bibfield  {author} {\bibinfo {author} {\bibfnamefont {Angelo}\ \bibnamefont
  {Bassi}}, \bibinfo {author} {\bibfnamefont {Kinjalk}\ \bibnamefont {Lochan}},
  \bibinfo {author} {\bibfnamefont {Seema}\ \bibnamefont {Satin}}, \bibinfo
  {author} {\bibfnamefont {Tejinder~P}\ \bibnamefont {Singh}}, \ and\ \bibinfo
  {author} {\bibfnamefont {Hendrik}\ \bibnamefont {Ulbricht}},\ }\bibfield
  {title} {\enquote {\bibinfo {title} {Models of wave-function collapse,
  underlying theories, and experimental tests},}\ }\href
  {https://doi.org/10.1103/RevModPhys.85.471} {\bibfield  {journal} {\bibinfo
  {journal} {Reviews of Modern Physics}\ }\textbf {\bibinfo {volume} {85}},\
  \bibinfo {pages} {471} (\bibinfo {year} {2013})}\BibitemShut {NoStop}%
\bibitem [{\citenamefont {Carlesso}\ \emph {et~al.}(2022)\citenamefont
  {Carlesso}, \citenamefont {Donadi}, \citenamefont {Ferialdi}, \citenamefont
  {Paternostro}, \citenamefont {Ulbricht},\ and\ \citenamefont
  {Bassi}}]{carlesso2022present}%
  \BibitemOpen
  \bibfield  {author} {\bibinfo {author} {\bibfnamefont {Matteo}\ \bibnamefont
  {Carlesso}}, \bibinfo {author} {\bibfnamefont {Sandro}\ \bibnamefont
  {Donadi}}, \bibinfo {author} {\bibfnamefont {Luca}\ \bibnamefont {Ferialdi}},
  \bibinfo {author} {\bibfnamefont {Mauro}\ \bibnamefont {Paternostro}},
  \bibinfo {author} {\bibfnamefont {Hendrik}\ \bibnamefont {Ulbricht}}, \ and\
  \bibinfo {author} {\bibfnamefont {Angelo}\ \bibnamefont {Bassi}},\ }\bibfield
   {title} {\enquote {\bibinfo {title} {Present status and future challenges of
  non-interferometric tests of collapse models},}\ }\href
  {https://doi.org/10.1038/s41567-021-01489-5} {\bibfield  {journal} {\bibinfo
  {journal} {Nature Physics}\ }\textbf {\bibinfo {volume} {18}},\ \bibinfo
  {pages} {243--250} (\bibinfo {year} {2022})}\BibitemShut {NoStop}%
\bibitem [{\citenamefont {Marchese}\ \emph {et~al.}(2021)\citenamefont
  {Marchese}, \citenamefont {Belenchia}, \citenamefont {Pirandola},\ and\
  \citenamefont {Paternostro}}]{Marchese1}%
  \BibitemOpen
  \bibfield  {author} {\bibinfo {author} {\bibfnamefont {M~M}\ \bibnamefont
  {Marchese}}, \bibinfo {author} {\bibfnamefont {A}~\bibnamefont {Belenchia}},
  \bibinfo {author} {\bibfnamefont {S}~\bibnamefont {Pirandola}}, \ and\
  \bibinfo {author} {\bibfnamefont {M}~\bibnamefont {Paternostro}},\ }\bibfield
   {title} {\enquote {\bibinfo {title} {An optomechanical platform for quantum
  hypothesis testing for collapse models},}\ }\href
  {https://doi.org/10.1088/1367-2630/abec0d} {\bibfield  {journal} {\bibinfo
  {journal} {New Journal of Physics}\ }\textbf {\bibinfo {volume} {23}},\
  \bibinfo {pages} {043022} (\bibinfo {year} {2021})}\BibitemShut {NoStop}%
\bibitem [{\citenamefont {Marchese}\ \emph {et~al.}(2023)\citenamefont
  {Marchese}, \citenamefont {Belenchia},\ and\ \citenamefont
  {Paternostro}}]{Marchese2}%
  \BibitemOpen
  \bibfield  {author} {\bibinfo {author} {\bibfnamefont {M~M}\ \bibnamefont
  {Marchese}}, \bibinfo {author} {\bibfnamefont {A}~\bibnamefont {Belenchia}},
  \ and\ \bibinfo {author} {\bibfnamefont {M}~\bibnamefont {Paternostro}},\
  }\bibfield  {title} {\enquote {\bibinfo {title} {Optomechanics-based quantum
  estimation theory for collapse models},}\ }\href
  {https://doi.org/10.3390/e25030500} {\bibfield  {journal} {\bibinfo
  {journal} {Entropy}\ }\textbf {\bibinfo {volume} {25}},\ \bibinfo {pages}
  {500} (\bibinfo {year} {2023})}\BibitemShut {NoStop}%
\bibitem [{\citenamefont {Smirne}\ \emph {et~al.}(2013)\citenamefont {Smirne},
  \citenamefont {Cialdi}, \citenamefont {Anelli}, \citenamefont {Paris},\ and\
  \citenamefont {Vacchini}}]{Smirne:2013}%
  \BibitemOpen
  \bibfield  {author} {\bibinfo {author} {\bibfnamefont {Andrea}\ \bibnamefont
  {Smirne}}, \bibinfo {author} {\bibfnamefont {Simone}\ \bibnamefont {Cialdi}},
  \bibinfo {author} {\bibfnamefont {Giorgio}\ \bibnamefont {Anelli}}, \bibinfo
  {author} {\bibfnamefont {Matteo G.~A.}\ \bibnamefont {Paris}}, \ and\
  \bibinfo {author} {\bibfnamefont {Bassano}\ \bibnamefont {Vacchini}},\
  }\bibfield  {title} {\enquote {\bibinfo {title} {Quantum probes to
  experimentally assess correlations in a composite system},}\ }\href {\doibase
  10.1103/PhysRevA.88.012108} {\bibfield  {journal} {\bibinfo  {journal} {Phys.
  Rev. A}\ }\textbf {\bibinfo {volume} {88}},\ \bibinfo {pages} {012108}
  (\bibinfo {year} {2013})}\BibitemShut {NoStop}%
\bibitem [{\citenamefont {Benedetti}\ \emph {et~al.}(2018)\citenamefont
  {Benedetti}, \citenamefont {Salari~Sehdaran}, \citenamefont {Zandi},\ and\
  \citenamefont {Paris}}]{Benedetti2018}%
  \BibitemOpen
  \bibfield  {author} {\bibinfo {author} {\bibfnamefont {Claudia}\ \bibnamefont
  {Benedetti}}, \bibinfo {author} {\bibfnamefont {Fahimeh}\ \bibnamefont
  {Salari~Sehdaran}}, \bibinfo {author} {\bibfnamefont {Mohammad~H.}\
  \bibnamefont {Zandi}}, \ and\ \bibinfo {author} {\bibfnamefont {Matteo
  G.~A.}\ \bibnamefont {Paris}},\ }\bibfield  {title} {\enquote {\bibinfo
  {title} {Quantum probes for the cutoff frequency of {O}hmic environments},}\
  }\href {\doibase 10.1103/PhysRevA.97.012126} {\bibfield  {journal} {\bibinfo
  {journal} {Phys. Rev. A}\ }\textbf {\bibinfo {volume} {97}},\ \bibinfo
  {pages} {012126} (\bibinfo {year} {2018})}\BibitemShut {NoStop}%
\bibitem [{\citenamefont {Tamascelli}\ \emph {et~al.}(2020)\citenamefont
  {Tamascelli}, \citenamefont {Benedetti}, \citenamefont {Breuer},\ and\
  \citenamefont {Paris}}]{Tamascelli2020}%
  \BibitemOpen
  \bibfield  {author} {\bibinfo {author} {\bibfnamefont {Dario}\ \bibnamefont
  {Tamascelli}}, \bibinfo {author} {\bibfnamefont {Claudia}\ \bibnamefont
  {Benedetti}}, \bibinfo {author} {\bibfnamefont {Heinz-Peter}\ \bibnamefont
  {Breuer}}, \ and\ \bibinfo {author} {\bibfnamefont {Matteo G~A}\ \bibnamefont
  {Paris}},\ }\bibfield  {title} {\enquote {\bibinfo {title} {Quantum probing
  beyond pure dephasing},}\ }\href {\doibase 10.1088/1367-2630/aba0e5}
  {\bibfield  {journal} {\bibinfo  {journal} {New Journal of Physics}\ }\textbf
  {\bibinfo {volume} {22}},\ \bibinfo {pages} {083027} (\bibinfo {year}
  {2020})}\BibitemShut {NoStop}%
\bibitem [{\citenamefont {Mehboudi}\ \emph {et~al.}(2019)\citenamefont
  {Mehboudi}, \citenamefont {Sanpera},\ and\ \citenamefont
  {Correa}}]{Mehboudi2019}%
  \BibitemOpen
  \bibfield  {author} {\bibinfo {author} {\bibfnamefont {Mohammad}\
  \bibnamefont {Mehboudi}}, \bibinfo {author} {\bibfnamefont {Anna}\
  \bibnamefont {Sanpera}}, \ and\ \bibinfo {author} {\bibfnamefont {Luis~A}\
  \bibnamefont {Correa}},\ }\bibfield  {title} {\enquote {\bibinfo {title}
  {Thermometry in the quantum regime: recent theoretical progress},}\ }\href
  {\doibase 10.1088/1751-8121/ab2828} {\bibfield  {journal} {\bibinfo
  {journal} {Journal of Physics A: Mathematical and Theoretical}\ }\textbf
  {\bibinfo {volume} {52}},\ \bibinfo {pages} {303001} (\bibinfo {year}
  {2019})}\BibitemShut {NoStop}%
\bibitem [{\citenamefont {Mascherpa}\ \emph {et~al.}(2017)\citenamefont
  {Mascherpa}, \citenamefont {Smirne}, \citenamefont {Huelga},\ and\
  \citenamefont {Plenio}}]{Mascherpa:2017}%
  \BibitemOpen
  \bibfield  {author} {\bibinfo {author} {\bibfnamefont {F.}~\bibnamefont
  {Mascherpa}}, \bibinfo {author} {\bibfnamefont {A.}~\bibnamefont {Smirne}},
  \bibinfo {author} {\bibfnamefont {S.~F.}\ \bibnamefont {Huelga}}, \ and\
  \bibinfo {author} {\bibfnamefont {M.~B.}\ \bibnamefont {Plenio}},\ }\bibfield
   {title} {\enquote {\bibinfo {title} {Open systems with error bounds:
  Spin-boson model with spectral density variations},}\ }\href {\doibase
  10.1103/PhysRevLett.118.100401} {\bibfield  {journal} {\bibinfo  {journal}
  {Phys. Rev. Lett.}\ }\textbf {\bibinfo {volume} {118}},\ \bibinfo {pages}
  {100401} (\bibinfo {year} {2017})}\BibitemShut {NoStop}%
\bibitem [{\citenamefont {Konopik}\ and\ \citenamefont
  {Lutz}(2019)}]{Konopik:2019}%
  \BibitemOpen
  \bibfield  {author} {\bibinfo {author} {\bibfnamefont {Michael}\ \bibnamefont
  {Konopik}}\ and\ \bibinfo {author} {\bibfnamefont {Eric}\ \bibnamefont
  {Lutz}},\ }\bibfield  {title} {\enquote {\bibinfo {title} {Quantum response
  theory for nonequilibrium steady states},}\ }\href {\doibase
  10.1103/PhysRevResearch.1.033156} {\bibfield  {journal} {\bibinfo  {journal}
  {Phys. Rev. Res.}\ }\textbf {\bibinfo {volume} {1}},\ \bibinfo {pages}
  {033156} (\bibinfo {year} {2019})}\BibitemShut {NoStop}%
\bibitem [{\citenamefont {Blair}\ \emph {et~al.}(2024)\citenamefont {Blair},
  \citenamefont {Zicari}, \citenamefont {Belenchia}, \citenamefont {Ferraro},\
  and\ \citenamefont {Paternostro}}]{Blair:2024}%
  \BibitemOpen
  \bibfield  {author} {\bibinfo {author} {\bibfnamefont {S.}~\bibnamefont
  {Blair}}, \bibinfo {author} {\bibfnamefont {G.}~\bibnamefont {Zicari}},
  \bibinfo {author} {\bibfnamefont {A.}~\bibnamefont {Belenchia}}, \bibinfo
  {author} {\bibfnamefont {A.}~\bibnamefont {Ferraro}}, \ and\ \bibinfo
  {author} {\bibfnamefont {M.}~\bibnamefont {Paternostro}},\ }\bibfield
  {title} {\enquote {\bibinfo {title} {Nonequilibrium quantum probing through
  linear response},}\ }\href {\doibase 10.1103/PhysRevResearch.6.013152}
  {\bibfield  {journal} {\bibinfo  {journal} {Phys. Rev. Res.}\ }\textbf
  {\bibinfo {volume} {6}},\ \bibinfo {pages} {013152} (\bibinfo {year}
  {2024})}\BibitemShut {NoStop}%
\bibitem [{\citenamefont {Degen}\ \emph {et~al.}(2017)\citenamefont {Degen},
  \citenamefont {Reinhard},\ and\ \citenamefont {Cappellaro}}]{Degen2017}%
  \BibitemOpen
  \bibfield  {author} {\bibinfo {author} {\bibfnamefont {C.~L.}\ \bibnamefont
  {Degen}}, \bibinfo {author} {\bibfnamefont {F.}~\bibnamefont {Reinhard}}, \
  and\ \bibinfo {author} {\bibfnamefont {P.}~\bibnamefont {Cappellaro}},\
  }\bibfield  {title} {\enquote {\bibinfo {title} {Quantum sensing},}\ }\href
  {\doibase 10.1103/RevModPhys.89.035002} {\bibfield  {journal} {\bibinfo
  {journal} {Rev. Mod. Phys.}\ }\textbf {\bibinfo {volume} {89}},\ \bibinfo
  {pages} {035002} (\bibinfo {year} {2017})}\BibitemShut {NoStop}%
\bibitem [{\citenamefont {Demkowicz-Dobrza\'{n}ski}\ \emph
  {et~al.}(2012)\citenamefont {Demkowicz-Dobrza\'{n}ski}, \citenamefont
  {Ko{\l}ody\'{n}ski},\ and\ \citenamefont {Gu\c{t}\u{a}}}]{Demkowicz2012}%
  \BibitemOpen
  \bibfield  {author} {\bibinfo {author} {\bibfnamefont {Rafa\l}\ \bibnamefont
  {Demkowicz-Dobrza\'{n}ski}}, \bibinfo {author} {\bibfnamefont {Jan}\
  \bibnamefont {Ko{\l}ody\'{n}ski}}, \ and\ \bibinfo {author} {\bibfnamefont
  {M\u{a}d\u{a}lin}\ \bibnamefont {Gu\c{t}\u{a}}},\ }\bibfield  {title}
  {\enquote {\bibinfo {title} {The elusive {H}eisenberg limit in
  quantum-enhanced metrologys},}\ }\href {\doibase 10.1038/ncomms2067}
  {\bibfield  {journal} {\bibinfo  {journal} {Nature Communications}\ }\textbf
  {\bibinfo {volume} {3}} (\bibinfo {year} {2012}),\
  10.1038/ncomms2067}\BibitemShut {NoStop}%
\bibitem [{\citenamefont {Leibfried}\ \emph {et~al.}(2004)\citenamefont
  {Leibfried}, \citenamefont {Barrett}, \citenamefont {Schaetz}, \citenamefont
  {Britton}, \citenamefont {Chiaverini}, \citenamefont {Itano}, \citenamefont
  {Jost}, \citenamefont {Langer},\ and\ \citenamefont
  {Wineland}}]{Leibfried2004}%
  \BibitemOpen
  \bibfield  {author} {\bibinfo {author} {\bibfnamefont {D.}~\bibnamefont
  {Leibfried}}, \bibinfo {author} {\bibfnamefont {M.~D.}\ \bibnamefont
  {Barrett}}, \bibinfo {author} {\bibfnamefont {T.}~\bibnamefont {Schaetz}},
  \bibinfo {author} {\bibfnamefont {J.}~\bibnamefont {Britton}}, \bibinfo
  {author} {\bibfnamefont {J.}~\bibnamefont {Chiaverini}}, \bibinfo {author}
  {\bibfnamefont {W.~M.}\ \bibnamefont {Itano}}, \bibinfo {author}
  {\bibfnamefont {J.~D.}\ \bibnamefont {Jost}}, \bibinfo {author}
  {\bibfnamefont {C.}~\bibnamefont {Langer}}, \ and\ \bibinfo {author}
  {\bibfnamefont {D.~J.}\ \bibnamefont {Wineland}},\ }\bibfield  {title}
  {\enquote {\bibinfo {title} {Toward {H}eisenberg-limited spectroscopy with
  multiparticle entangled states},}\ }\href {\doibase 10.1126/science.1097576}
  {\bibfield  {journal} {\bibinfo  {journal} {Science}\ }\textbf {\bibinfo
  {volume} {304}},\ \bibinfo {pages} {1476--1478} (\bibinfo {year}
  {2004})}\BibitemShut {NoStop}%
\bibitem [{\citenamefont {Giovannetti}\ \emph {et~al.}(2004)\citenamefont
  {Giovannetti}, \citenamefont {Lloyd},\ and\ \citenamefont
  {Maccone}}]{Giovannetti2004}%
  \BibitemOpen
  \bibfield  {author} {\bibinfo {author} {\bibfnamefont {V}~\bibnamefont
  {Giovannetti}}, \bibinfo {author} {\bibfnamefont {S}~\bibnamefont {Lloyd}}, \
  and\ \bibinfo {author} {\bibfnamefont {L}~\bibnamefont {Maccone}},\
  }\bibfield  {title} {\enquote {\bibinfo {title} {Quantum-enhanced
  measurements: Beating the standard quantum limit},}\ }\href
  {https://doi.org/10.1126/science.1104149} {\bibfield  {journal} {\bibinfo
  {journal} {Science}\ }\textbf {\bibinfo {volume} {306}},\ \bibinfo {pages}
  {1330} (\bibinfo {year} {2004})}\BibitemShut {NoStop}%
\bibitem [{\citenamefont {Campos~Venuti}\ and\ \citenamefont
  {Zanardi}(2007)}]{CamposVenuti:2007}%
  \BibitemOpen
  \bibfield  {author} {\bibinfo {author} {\bibfnamefont {Lorenzo}\ \bibnamefont
  {Campos~Venuti}}\ and\ \bibinfo {author} {\bibfnamefont {Paolo}\ \bibnamefont
  {Zanardi}},\ }\bibfield  {title} {\enquote {\bibinfo {title} {Quantum
  critical scaling of the geometric tensors},}\ }\href {\doibase
  10.1103/PhysRevLett.99.095701} {\bibfield  {journal} {\bibinfo  {journal}
  {Phys. Rev. Lett.}\ }\textbf {\bibinfo {volume} {99}},\ \bibinfo {pages}
  {095701} (\bibinfo {year} {2007})}\BibitemShut {NoStop}%
\bibitem [{\citenamefont {Zanardi}\ \emph {et~al.}(2008)\citenamefont
  {Zanardi}, \citenamefont {Paris},\ and\ \citenamefont
  {Campos~Venuti}}]{Zanardi:2008}%
  \BibitemOpen
  \bibfield  {author} {\bibinfo {author} {\bibfnamefont {Paolo}\ \bibnamefont
  {Zanardi}}, \bibinfo {author} {\bibfnamefont {Matteo G.~A.}\ \bibnamefont
  {Paris}}, \ and\ \bibinfo {author} {\bibfnamefont {Lorenzo}\ \bibnamefont
  {Campos~Venuti}},\ }\bibfield  {title} {\enquote {\bibinfo {title} {Quantum
  criticality as a resource for quantum estimation},}\ }\href {\doibase
  10.1103/PhysRevA.78.042105} {\bibfield  {journal} {\bibinfo  {journal} {Phys.
  Rev. A}\ }\textbf {\bibinfo {volume} {78}},\ \bibinfo {pages} {042105}
  (\bibinfo {year} {2008})}\BibitemShut {NoStop}%
\bibitem [{\citenamefont {Fr\'erot}\ and\ \citenamefont
  {Roscilde}(2018)}]{Frerot:2018}%
  \BibitemOpen
  \bibfield  {author} {\bibinfo {author} {\bibfnamefont {Ir\'en\'ee}\
  \bibnamefont {Fr\'erot}}\ and\ \bibinfo {author} {\bibfnamefont {Tommaso}\
  \bibnamefont {Roscilde}},\ }\bibfield  {title} {\enquote {\bibinfo {title}
  {Quantum critical metrology},}\ }\href {\doibase
  10.1103/PhysRevLett.121.020402} {\bibfield  {journal} {\bibinfo  {journal}
  {Phys. Rev. Lett.}\ }\textbf {\bibinfo {volume} {121}},\ \bibinfo {pages}
  {020402} (\bibinfo {year} {2018})}\BibitemShut {NoStop}%
\bibitem [{\citenamefont {Rams}\ \emph {et~al.}(2018)\citenamefont {Rams},
  \citenamefont {Sierant}, \citenamefont {Dutta}, \citenamefont {Horodecki},\
  and\ \citenamefont {Zakrzewski}}]{Rams:2018}%
  \BibitemOpen
  \bibfield  {author} {\bibinfo {author} {\bibfnamefont {Marek~M.}\
  \bibnamefont {Rams}}, \bibinfo {author} {\bibfnamefont {Piotr}\ \bibnamefont
  {Sierant}}, \bibinfo {author} {\bibfnamefont {Omyoti}\ \bibnamefont {Dutta}},
  \bibinfo {author} {\bibfnamefont {Pawe\l{}}\ \bibnamefont {Horodecki}}, \
  and\ \bibinfo {author} {\bibfnamefont {Jakub}\ \bibnamefont {Zakrzewski}},\
  }\bibfield  {title} {\enquote {\bibinfo {title} {At the limits of
  criticality-based quantum metrology: Apparent super-{H}eisenberg scaling
  revisited},}\ }\href {\doibase 10.1103/PhysRevX.8.021022} {\bibfield
  {journal} {\bibinfo  {journal} {Phys. Rev. X}\ }\textbf {\bibinfo {volume}
  {8}},\ \bibinfo {pages} {021022} (\bibinfo {year} {2018})}\BibitemShut
  {NoStop}%
\bibitem [{\citenamefont {Hotter}\ \emph {et~al.}(2024)\citenamefont {Hotter},
  \citenamefont {Ritsch},\ and\ \citenamefont {Gietka}}]{Hotter:2024}%
  \BibitemOpen
  \bibfield  {author} {\bibinfo {author} {\bibfnamefont {Christoph}\
  \bibnamefont {Hotter}}, \bibinfo {author} {\bibfnamefont {Helmut}\
  \bibnamefont {Ritsch}}, \ and\ \bibinfo {author} {\bibfnamefont {Karol}\
  \bibnamefont {Gietka}},\ }\bibfield  {title} {\enquote {\bibinfo {title}
  {Combining critical and quantum metrology},}\ }\href {\doibase
  10.1103/PhysRevLett.132.060801} {\bibfield  {journal} {\bibinfo  {journal}
  {Phys. Rev. Lett.}\ }\textbf {\bibinfo {volume} {132}},\ \bibinfo {pages}
  {060801} (\bibinfo {year} {2024})}\BibitemShut {NoStop}%
\bibitem [{\citenamefont {Mihailescu}\ \emph {et~al.}(2024)\citenamefont
  {Mihailescu}, \citenamefont {Bayat}, \citenamefont {Campbell},\ and\
  \citenamefont {Mitchell}}]{Mihailescu:2024}%
  \BibitemOpen
  \bibfield  {author} {\bibinfo {author} {\bibfnamefont {George}\ \bibnamefont
  {Mihailescu}}, \bibinfo {author} {\bibfnamefont {Abolfazl}\ \bibnamefont
  {Bayat}}, \bibinfo {author} {\bibfnamefont {Steve}\ \bibnamefont {Campbell}},
  \ and\ \bibinfo {author} {\bibfnamefont {Andrew~K}\ \bibnamefont
  {Mitchell}},\ }\bibfield  {title} {\enquote {\bibinfo {title} {Multiparameter
  critical quantum metrology with impurity probes},}\ }\href {\doibase
  10.1088/2058-9565/ad438d} {\bibfield  {journal} {\bibinfo  {journal} {Quantum
  Science and Technology}\ }\textbf {\bibinfo {volume} {9}},\ \bibinfo {pages}
  {035033} (\bibinfo {year} {2024})}\BibitemShut {NoStop}%
\bibitem [{\citenamefont {Garbe}\ \emph {et~al.}(2020)\citenamefont {Garbe},
  \citenamefont {Bina}, \citenamefont {Keller}, \citenamefont {Paris},\ and\
  \citenamefont {Felicetti}}]{Garbe:2020}%
  \BibitemOpen
  \bibfield  {author} {\bibinfo {author} {\bibfnamefont {Louis}\ \bibnamefont
  {Garbe}}, \bibinfo {author} {\bibfnamefont {Matteo}\ \bibnamefont {Bina}},
  \bibinfo {author} {\bibfnamefont {Arne}\ \bibnamefont {Keller}}, \bibinfo
  {author} {\bibfnamefont {Matteo G.~A.}\ \bibnamefont {Paris}}, \ and\
  \bibinfo {author} {\bibfnamefont {Simone}\ \bibnamefont {Felicetti}},\
  }\bibfield  {title} {\enquote {\bibinfo {title} {Critical quantum metrology
  with a finite-component quantum phase transition},}\ }\href {\doibase
  10.1103/PhysRevLett.124.120504} {\bibfield  {journal} {\bibinfo  {journal}
  {Phys. Rev. Lett.}\ }\textbf {\bibinfo {volume} {124}},\ \bibinfo {pages}
  {120504} (\bibinfo {year} {2020})}\BibitemShut {NoStop}%
\bibitem [{\citenamefont {Garbe}\ \emph {et~al.}(2022)\citenamefont {Garbe},
  \citenamefont {Abah}, \citenamefont {Felicetti},\ and\ \citenamefont
  {Puebla}}]{Garbe:2022}%
  \BibitemOpen
  \bibfield  {author} {\bibinfo {author} {\bibfnamefont {Louis}\ \bibnamefont
  {Garbe}}, \bibinfo {author} {\bibfnamefont {Obinna}\ \bibnamefont {Abah}},
  \bibinfo {author} {\bibfnamefont {Simone}\ \bibnamefont {Felicetti}}, \ and\
  \bibinfo {author} {\bibfnamefont {Ricardo}\ \bibnamefont {Puebla}},\
  }\bibfield  {title} {\enquote {\bibinfo {title} {Critical quantum metrology
  with fully-connected models: from {H}eisenberg to {K}ibble--{Z}urek
  scaling},}\ }\href {\doibase 10.1088/2058-9565/ac6ca5} {\bibfield  {journal}
  {\bibinfo  {journal} {Quantum Science and Technology}\ }\textbf {\bibinfo
  {volume} {7}},\ \bibinfo {pages} {035010} (\bibinfo {year}
  {2022})}\BibitemShut {NoStop}%
\bibitem [{\citenamefont {Di~Candia}\ \emph {et~al.}(2023)\citenamefont
  {Di~Candia}, \citenamefont {Minganti}, \citenamefont {Petrovnin},
  \citenamefont {Paraoanu},\ and\ \citenamefont {Felicetti}}]{DiCandia:2023}%
  \BibitemOpen
  \bibfield  {author} {\bibinfo {author} {\bibfnamefont {R.}~\bibnamefont
  {Di~Candia}}, \bibinfo {author} {\bibfnamefont {F.}~\bibnamefont {Minganti}},
  \bibinfo {author} {\bibfnamefont {K.~V.}\ \bibnamefont {Petrovnin}}, \bibinfo
  {author} {\bibfnamefont {G.~S.}\ \bibnamefont {Paraoanu}}, \ and\ \bibinfo
  {author} {\bibfnamefont {S.}~\bibnamefont {Felicetti}},\ }\bibfield  {title}
  {\enquote {\bibinfo {title} {Critical parametric quantum sensing},}\ }\href
  {\doibase 10.1038/s41534-023-00690-z} {\bibfield  {journal} {\bibinfo
  {journal} {npj Quantum Information}\ }\textbf {\bibinfo {volume} {9}},\
  \bibinfo {pages} {23} (\bibinfo {year} {2023})}\BibitemShut {NoStop}%
\bibitem [{\citenamefont {Alushi}\ \emph {et~al.}(2024)\citenamefont {Alushi},
  \citenamefont {G\'orecki}, \citenamefont {Felicetti},\ and\ \citenamefont
  {Di~Candia}}]{Alushi:2024}%
  \BibitemOpen
  \bibfield  {author} {\bibinfo {author} {\bibfnamefont {U.}~\bibnamefont
  {Alushi}}, \bibinfo {author} {\bibfnamefont {W.}~\bibnamefont {G\'orecki}},
  \bibinfo {author} {\bibfnamefont {S.}~\bibnamefont {Felicetti}}, \ and\
  \bibinfo {author} {\bibfnamefont {R.}~\bibnamefont {Di~Candia}},\ }\bibfield
  {title} {\enquote {\bibinfo {title} {Optimality and noise resilience of
  critical quantum sensing},}\ }\href {\doibase 10.1103/PhysRevLett.133.040801}
  {\bibfield  {journal} {\bibinfo  {journal} {Phys. Rev. Lett.}\ }\textbf
  {\bibinfo {volume} {133}},\ \bibinfo {pages} {040801} (\bibinfo {year}
  {2024})}\BibitemShut {NoStop}%
\bibitem [{\citenamefont {Vacchini}(2007)}]{Vacchini_2007}%
  \BibitemOpen
  \bibfield  {author} {\bibinfo {author} {\bibfnamefont {Bassano}\ \bibnamefont
  {Vacchini}},\ }\bibfield  {title} {\enquote {\bibinfo {title} {On the precise
  connection between the {GRW} master equation and master equations for the
  description of decoherence},}\ }\href {\doibase 10.1088/1751-8113/40/10/015}
  {\bibfield  {journal} {\bibinfo  {journal} {Journal of Physics A:
  Mathematical and Theoretical}\ }\textbf {\bibinfo {volume} {40}},\ \bibinfo
  {pages} {2463} (\bibinfo {year} {2007})}\BibitemShut {NoStop}%
\bibitem [{\citenamefont {Breuer}\ and\ \citenamefont
  {Petruccione}(2002)}]{Breuer-Petruccione}%
  \BibitemOpen
  \bibfield  {author} {\bibinfo {author} {\bibfnamefont {Heinz-Peter}\
  \bibnamefont {Breuer}}\ and\ \bibinfo {author} {\bibfnamefont {Francesco}\
  \bibnamefont {Petruccione}},\ }\href {\doibase
  10.1093/acprof:oso/9780199213900.001.0001} {\emph {\bibinfo {title} {The
  Theory of Open Quantum Systems}}}\ (\bibinfo  {publisher} {Oxford University
  Press},\ \bibinfo {address} {Oxford},\ \bibinfo {year} {2002})\BibitemShut
  {NoStop}%
\bibitem [{\citenamefont {Vacchini}(2024)}]{Vacchini:2024}%
  \BibitemOpen
  \bibfield  {author} {\bibinfo {author} {\bibfnamefont {Bassano}\ \bibnamefont
  {Vacchini}},\ }\href {\doibase 10.1007/978-3-031-58218-9} {\emph {\bibinfo
  {title} {Open Quantum Systems}}},\ \bibinfo {edition} {1st}\ ed.,\ Graduate
  Texts in Physics\ (\bibinfo  {publisher} {Springer Cham},\ \bibinfo {year}
  {2024})\BibitemShut {NoStop}%
\bibitem [{\citenamefont {Haase}\ \emph {et~al.}(2016)\citenamefont {Haase},
  \citenamefont {Smirne}, \citenamefont {Huelga}, \citenamefont
  {Ko{\l}odynski},\ and\ \citenamefont
  {Demkowicz-Dobrza\'{n}ski}}]{Haase:2016}%
  \BibitemOpen
  \bibfield  {author} {\bibinfo {author} {\bibfnamefont {J.~F.}\ \bibnamefont
  {Haase}}, \bibinfo {author} {\bibfnamefont {A.}~\bibnamefont {Smirne}},
  \bibinfo {author} {\bibfnamefont {S.~F.}\ \bibnamefont {Huelga}}, \bibinfo
  {author} {\bibfnamefont {J.}~\bibnamefont {Ko{\l}odynski}}, \ and\ \bibinfo
  {author} {\bibfnamefont {R.}~\bibnamefont {Demkowicz-Dobrza\'{n}ski}},\
  }\bibfield  {title} {\enquote {\bibinfo {title} {Precision limits in quantum
  metrology with open quantum systems},}\ }\href {\doibase
  doi:10.1515/qmetro-2018-0002} {\bibfield  {journal} {\bibinfo  {journal}
  {Quantum Measurements and Quantum Metrology}\ }\textbf {\bibinfo {volume}
  {5}},\ \bibinfo {pages} {13--39} (\bibinfo {year} {2016})}\BibitemShut
  {NoStop}%
\bibitem [{\citenamefont {Fujiwara}(2001)}]{Fujiwara:2001}%
  \BibitemOpen
  \bibfield  {author} {\bibinfo {author} {\bibfnamefont {Akio}\ \bibnamefont
  {Fujiwara}},\ }\bibfield  {title} {\enquote {\bibinfo {title} {Quantum
  channel identification problem},}\ }\href {\doibase
  10.1103/PhysRevA.63.042304} {\bibfield  {journal} {\bibinfo  {journal} {Phys.
  Rev. A}\ }\textbf {\bibinfo {volume} {63}},\ \bibinfo {pages} {042304}
  (\bibinfo {year} {2001})}\BibitemShut {NoStop}%
\bibitem [{\citenamefont {Monras}\ and\ \citenamefont
  {Paris}(2007)}]{Monras:2007}%
  \BibitemOpen
  \bibfield  {author} {\bibinfo {author} {\bibfnamefont {Alex}\ \bibnamefont
  {Monras}}\ and\ \bibinfo {author} {\bibfnamefont {Matteo G.~A.}\ \bibnamefont
  {Paris}},\ }\bibfield  {title} {\enquote {\bibinfo {title} {Optimal quantum
  estimation of loss in bosonic channels},}\ }\href {\doibase
  10.1103/PhysRevLett.98.160401} {\bibfield  {journal} {\bibinfo  {journal}
  {Phys. Rev. Lett.}\ }\textbf {\bibinfo {volume} {98}},\ \bibinfo {pages}
  {160401} (\bibinfo {year} {2007})}\BibitemShut {NoStop}%
\bibitem [{\citenamefont {Ghirardi}(1999)}]{Ghirardi:1999}%
  \BibitemOpen
  \bibfield  {author} {\bibinfo {author} {\bibfnamefont {GianCarlo}\
  \bibnamefont {Ghirardi}},\ }\bibfield  {title} {\enquote {\bibinfo {title}
  {Quantum superpositions and definite perceptions: envisaging new feasible
  experimental tests},}\ }\href {\doibase
  https://doi.org/10.1016/S0375-9601(99)00646-5} {\bibfield  {journal}
  {\bibinfo  {journal} {Physics Letters A}\ }\textbf {\bibinfo {volume}
  {262}},\ \bibinfo {pages} {1--14} (\bibinfo {year} {1999})}\BibitemShut
  {NoStop}%
\bibitem [{\citenamefont {Bassi}\ and\ \citenamefont
  {Ippoliti}(2004)}]{Bassi:2004}%
  \BibitemOpen
  \bibfield  {author} {\bibinfo {author} {\bibfnamefont {Angelo}\ \bibnamefont
  {Bassi}}\ and\ \bibinfo {author} {\bibfnamefont {Emiliano}\ \bibnamefont
  {Ippoliti}},\ }\bibfield  {title} {\enquote {\bibinfo {title} {Numerical
  analysis of a spontaneous collapse model for a two-level system},}\ }\href
  {\doibase 10.1103/PhysRevA.69.012105} {\bibfield  {journal} {\bibinfo
  {journal} {Phys. Rev. A}\ }\textbf {\bibinfo {volume} {69}},\ \bibinfo
  {pages} {012105} (\bibinfo {year} {2004})}\BibitemShut {NoStop}%
\bibitem [{\citenamefont {Bahrami}\ \emph {et~al.}(2013)\citenamefont
  {Bahrami}, \citenamefont {Donadi}, \citenamefont {Ferialdi}, \citenamefont
  {Bassi}, \citenamefont {Curceanu}, \citenamefont {C.},\ and\ \citenamefont
  {A.}}]{Bahrami:2013}%
  \BibitemOpen
  \bibfield  {author} {\bibinfo {author} {\bibfnamefont {M.}~\bibnamefont
  {Bahrami}}, \bibinfo {author} {\bibfnamefont {S.}~\bibnamefont {Donadi}},
  \bibinfo {author} {\bibfnamefont {L.}~\bibnamefont {Ferialdi}}, \bibinfo
  {author} {\bibfnamefont {A.}~\bibnamefont {Bassi}}, \bibinfo {author}
  {\bibnamefont {Curceanu}}, \bibinfo {author} {\bibfnamefont {Di~Domenico}\
  \bibnamefont {C.}}, \ and\ \bibinfo {author} {\bibfnamefont {B.~C.}\
  \bibnamefont {A.}, \bibfnamefont {Hiesmayr}},\ }\bibfield  {title} {\enquote
  {\bibinfo {title} {Are collapse models testable with quantum oscillating
  systems? {T}he case of neutrinos, kaons, chiral molecules},}\ }\href
  {\doibase 10.1038/srep01952} {\bibfield  {journal} {\bibinfo  {journal}
  {Scientific Reports}\ }\textbf {\bibinfo {volume} {3}},\ \bibinfo {pages}
  {1952} (\bibinfo {year} {2013})}\BibitemShut {NoStop}%
\bibitem [{\citenamefont {Dziarmaga}(2005)}]{Dziamarga:2005}%
  \BibitemOpen
  \bibfield  {author} {\bibinfo {author} {\bibfnamefont {Jacek}\ \bibnamefont
  {Dziarmaga}},\ }\bibfield  {title} {\enquote {\bibinfo {title} {Dynamics of a
  quantum phase transition: Exact solution of the quantum {I}sing model},}\
  }\href {\doibase 10.1103/PhysRevLett.95.245701} {\bibfield  {journal}
  {\bibinfo  {journal} {Phys. Rev. Lett.}\ }\textbf {\bibinfo {volume} {95}},\
  \bibinfo {pages} {245701} (\bibinfo {year} {2005})}\BibitemShut {NoStop}%
\bibitem [{\citenamefont {Invernizzi}\ \emph {et~al.}(2008)\citenamefont
  {Invernizzi}, \citenamefont {Korbman}, \citenamefont {Campos~Venuti},\ and\
  \citenamefont {Paris}}]{Invernizzi:2008}%
  \BibitemOpen
  \bibfield  {author} {\bibinfo {author} {\bibfnamefont {Carmen}\ \bibnamefont
  {Invernizzi}}, \bibinfo {author} {\bibfnamefont {Michael}\ \bibnamefont
  {Korbman}}, \bibinfo {author} {\bibfnamefont {Lorenzo}\ \bibnamefont
  {Campos~Venuti}}, \ and\ \bibinfo {author} {\bibfnamefont {Matteo G.~A.}\
  \bibnamefont {Paris}},\ }\bibfield  {title} {\enquote {\bibinfo {title}
  {Optimal quantum estimation in spin systems at criticality},}\ }\href
  {\doibase 10.1103/PhysRevA.78.042106} {\bibfield  {journal} {\bibinfo
  {journal} {Phys. Rev. A}\ }\textbf {\bibinfo {volume} {78}},\ \bibinfo
  {pages} {042106} (\bibinfo {year} {2008})}\BibitemShut {NoStop}%
\bibitem [{\citenamefont {Seveso}\ and\ \citenamefont
  {Paris}(2020)}]{Seveso:2020}%
  \BibitemOpen
  \bibfield  {author} {\bibinfo {author} {\bibfnamefont {Luigi}\ \bibnamefont
  {Seveso}}\ and\ \bibinfo {author} {\bibfnamefont {Matteo G.~A.}\ \bibnamefont
  {Paris}},\ }\bibfield  {title} {\enquote {\bibinfo {title} {Quantum enhanced
  metrology of {H}amiltonian parameters beyond the {C}ram{\`e}r--{R}ao
  bound},}\ }\href {https://doi.org/10.1142/S0219749920300016} {\bibfield
  {journal} {\bibinfo  {journal} {International Journal of Quantum
  Information}\ }\textbf {\bibinfo {volume} {18}},\ \bibinfo {pages} {2030001}
  (\bibinfo {year} {2020})}\BibitemShut {NoStop}%
\bibitem [{\citenamefont {De~Pasquale}\ and\ \citenamefont
  {Stace}(2018)}]{DePasquale:2018}%
  \BibitemOpen
  \bibfield  {author} {\bibinfo {author} {\bibfnamefont {Antonella}\
  \bibnamefont {De~Pasquale}}\ and\ \bibinfo {author} {\bibfnamefont
  {Thomas~M.}\ \bibnamefont {Stace}},\ }\enquote {\bibinfo {title} {Quantum
  thermometry},}\ in\ \href {\doibase 10.1007/978-3-319-99046-0_21} {\emph
  {\bibinfo {booktitle} {Thermodynamics in the Quantum Regime: Fundamental
  Aspects and New Directions}}},\ \bibinfo {editor} {edited by\ \bibinfo
  {editor} {\bibfnamefont {Felix}\ \bibnamefont {Binder}}, \bibinfo {editor}
  {\bibfnamefont {Luis~A.}\ \bibnamefont {Correa}}, \bibinfo {editor}
  {\bibfnamefont {Christian}\ \bibnamefont {Gogolin}}, \bibinfo {editor}
  {\bibfnamefont {Janet}\ \bibnamefont {Anders}}, \ and\ \bibinfo {editor}
  {\bibfnamefont {Gerardo}\ \bibnamefont {Adesso}}}\ (\bibinfo  {publisher}
  {Springer International Publishing},\ \bibinfo {address} {Cham},\ \bibinfo
  {year} {2018})\ pp.\ \bibinfo {pages} {503--527}\BibitemShut {NoStop}%
\bibitem [{\citenamefont {Deffner}\ and\ \citenamefont
  {Campbell}(2019)}]{DeffnerCampbellBook}%
  \BibitemOpen
  \bibfield  {author} {\bibinfo {author} {\bibfnamefont {S.}~\bibnamefont
  {Deffner}}\ and\ \bibinfo {author} {\bibfnamefont {S.}~\bibnamefont
  {Campbell}},\ }\href {\doibase 10.1088/2053-2571/ab21c6} {\emph {\bibinfo
  {title} {Quantum Thermodynamics}}}\ (\bibinfo  {publisher} {Morgan \&
  Claypool Publishers},\ \bibinfo {year} {2019})\BibitemShut {NoStop}%
\bibitem [{\citenamefont {Correa}\ \emph {et~al.}(2015)\citenamefont {Correa},
  \citenamefont {Mehboudi}, \citenamefont {Adesso},\ and\ \citenamefont
  {Sanpera}}]{Correa:2015}%
  \BibitemOpen
  \bibfield  {author} {\bibinfo {author} {\bibfnamefont {Luis~A.}\ \bibnamefont
  {Correa}}, \bibinfo {author} {\bibfnamefont {Mohammad}\ \bibnamefont
  {Mehboudi}}, \bibinfo {author} {\bibfnamefont {Gerardo}\ \bibnamefont
  {Adesso}}, \ and\ \bibinfo {author} {\bibfnamefont {Anna}\ \bibnamefont
  {Sanpera}},\ }\bibfield  {title} {\enquote {\bibinfo {title} {Individual
  quantum probes for optimal thermometry},}\ }\href {\doibase
  10.1103/PhysRevLett.114.220405} {\bibfield  {journal} {\bibinfo  {journal}
  {Phys. Rev. Lett.}\ }\textbf {\bibinfo {volume} {114}},\ \bibinfo {pages}
  {220405} (\bibinfo {year} {2015})}\BibitemShut {NoStop}%
\bibitem [{\citenamefont {Mok}\ \emph {et~al.}(2021)\citenamefont {Mok},
  \citenamefont {Bharti}, \citenamefont {Kwek},\ and\ \citenamefont
  {Bayat}}]{Mok:2021}%
  \BibitemOpen
  \bibfield  {author} {\bibinfo {author} {\bibfnamefont {Wai-Keong}\
  \bibnamefont {Mok}}, \bibinfo {author} {\bibfnamefont {Kishor}\ \bibnamefont
  {Bharti}}, \bibinfo {author} {\bibfnamefont {Leong-Chuan}\ \bibnamefont
  {Kwek}}, \ and\ \bibinfo {author} {\bibfnamefont {Abolfazl}\ \bibnamefont
  {Bayat}},\ }\bibfield  {title} {\enquote {\bibinfo {title} {Optimal probes
  for global quantum thermometry},}\ }\href {\doibase
  10.1038/s42005-021-00572-w} {\bibfield  {journal} {\bibinfo  {journal}
  {Communications Physics}\ }\textbf {\bibinfo {volume} {4}},\ \bibinfo {pages}
  {1--8} (\bibinfo {year} {2021})}\BibitemShut {NoStop}%
\bibitem [{\citenamefont {Campbell}\ \emph {et~al.}(2018)\citenamefont
  {Campbell}, \citenamefont {Genoni},\ and\ \citenamefont
  {Deffner}}]{Campbell:2018}%
  \BibitemOpen
  \bibfield  {author} {\bibinfo {author} {\bibfnamefont {Steve}\ \bibnamefont
  {Campbell}}, \bibinfo {author} {\bibfnamefont {Marco~G}\ \bibnamefont
  {Genoni}}, \ and\ \bibinfo {author} {\bibfnamefont {Sebastian}\ \bibnamefont
  {Deffner}},\ }\bibfield  {title} {\enquote {\bibinfo {title} {Precision
  thermometry and the quantum speed limit},}\ }\href {\doibase
  10.1088/2058-9565/aaa641} {\bibfield  {journal} {\bibinfo  {journal} {Quantum
  Science and Technology}\ }\textbf {\bibinfo {volume} {3}},\ \bibinfo {pages}
  {025002} (\bibinfo {year} {2018})}\BibitemShut {NoStop}%
\bibitem [{\citenamefont {Brunelli}\ \emph {et~al.}(2011)\citenamefont
  {Brunelli}, \citenamefont {Olivares},\ and\ \citenamefont
  {Paris}}]{Brunelli:2011}%
  \BibitemOpen
  \bibfield  {author} {\bibinfo {author} {\bibfnamefont {Matteo}\ \bibnamefont
  {Brunelli}}, \bibinfo {author} {\bibfnamefont {Stefano}\ \bibnamefont
  {Olivares}}, \ and\ \bibinfo {author} {\bibfnamefont {Matteo G.~A.}\
  \bibnamefont {Paris}},\ }\bibfield  {title} {\enquote {\bibinfo {title}
  {Qubit thermometry for micromechanical resonators},}\ }\href {\doibase
  10.1103/PhysRevA.84.032105} {\bibfield  {journal} {\bibinfo  {journal} {Phys.
  Rev. A}\ }\textbf {\bibinfo {volume} {84}},\ \bibinfo {pages} {032105}
  (\bibinfo {year} {2011})}\BibitemShut {NoStop}%
\bibitem [{\citenamefont {Brunelli}\ \emph {et~al.}(2012)\citenamefont
  {Brunelli}, \citenamefont {Olivares}, \citenamefont {Paternostro},\ and\
  \citenamefont {Paris}}]{Brunelli:2012}%
  \BibitemOpen
  \bibfield  {author} {\bibinfo {author} {\bibfnamefont {Matteo}\ \bibnamefont
  {Brunelli}}, \bibinfo {author} {\bibfnamefont {Stefano}\ \bibnamefont
  {Olivares}}, \bibinfo {author} {\bibfnamefont {Mauro}\ \bibnamefont
  {Paternostro}}, \ and\ \bibinfo {author} {\bibfnamefont {Matteo G.~A.}\
  \bibnamefont {Paris}},\ }\bibfield  {title} {\enquote {\bibinfo {title}
  {Qubit-assisted thermometry of a quantum harmonic oscillator},}\ }\href
  {\doibase 10.1103/PhysRevA.86.012125} {\bibfield  {journal} {\bibinfo
  {journal} {Phys. Rev. A}\ }\textbf {\bibinfo {volume} {86}},\ \bibinfo
  {pages} {012125} (\bibinfo {year} {2012})}\BibitemShut {NoStop}%
\bibitem [{\citenamefont {Lindblad}(1976)}]{Lindblad:1976}%
  \BibitemOpen
  \bibfield  {author} {\bibinfo {author} {\bibfnamefont {G.}~\bibnamefont
  {Lindblad}},\ }\bibfield  {title} {\enquote {\bibinfo {title} {On the
  generators of quantum dynamical semigroups},}\ }\href {\doibase
  10.1007/BF01608499} {\bibfield  {journal} {\bibinfo  {journal}
  {Communications in Mathematical Physics}\ }\textbf {\bibinfo {volume} {48}},\
  \bibinfo {pages} {119--130} (\bibinfo {year} {1976})}\BibitemShut {NoStop}%
\bibitem [{\citenamefont {Gorini}\ \emph {et~al.}(1976)\citenamefont {Gorini},
  \citenamefont {Kossakowski},\ and\ \citenamefont {Sudarshan}}]{Gorini:1976}%
  \BibitemOpen
  \bibfield  {author} {\bibinfo {author} {\bibfnamefont {Vittorio}\
  \bibnamefont {Gorini}}, \bibinfo {author} {\bibfnamefont {Andrzej}\
  \bibnamefont {Kossakowski}}, \ and\ \bibinfo {author} {\bibfnamefont
  {E.~C.~G.}\ \bibnamefont {Sudarshan}},\ }\bibfield  {title} {\enquote
  {\bibinfo {title} {Completely positive dynamical semigroups of {$N$}-level
  systems},}\ }\href {\doibase 10.1063/1.522979} {\bibfield  {journal}
  {\bibinfo  {journal} {Journal of Mathematical Physics}\ }\textbf {\bibinfo
  {volume} {17}},\ \bibinfo {pages} {821--825} (\bibinfo {year}
  {1976})}\BibitemShut {NoStop}%
\bibitem [{\citenamefont {Gorini}\ \emph {et~al.}(1978)\citenamefont {Gorini},
  \citenamefont {Frigerio}, \citenamefont {Verri}, \citenamefont
  {Kossakowski},\ and\ \citenamefont {Sudarshan}}]{Gorini:1978}%
  \BibitemOpen
  \bibfield  {author} {\bibinfo {author} {\bibfnamefont {Vittorio}\
  \bibnamefont {Gorini}}, \bibinfo {author} {\bibfnamefont {Alberto}\
  \bibnamefont {Frigerio}}, \bibinfo {author} {\bibfnamefont {Maurizio}\
  \bibnamefont {Verri}}, \bibinfo {author} {\bibfnamefont {Andrzej}\
  \bibnamefont {Kossakowski}}, \ and\ \bibinfo {author} {\bibfnamefont
  {E.C.G.}\ \bibnamefont {Sudarshan}},\ }\bibfield  {title} {\enquote {\bibinfo
  {title} {Properties of quantum {M}arkovian master equations},}\ }\href
  {\doibase 10.1016/0034-4877(78)90050-2} {\bibfield  {journal} {\bibinfo
  {journal} {Reports on Mathematical Physics}\ }\textbf {\bibinfo {volume}
  {13}},\ \bibinfo {pages} {149--173} (\bibinfo {year} {1978})}\BibitemShut
  {NoStop}%
\bibitem [{\citenamefont {Paris}(2009)}]{Paris:2009}%
  \BibitemOpen
  \bibfield  {author} {\bibinfo {author} {\bibfnamefont {Matteo G.~A.}\
  \bibnamefont {Paris}},\ }\bibfield  {title} {\enquote {\bibinfo {title}
  {Quantum estimation for quantum technology},}\ }\href {\doibase
  10.1142/S0219749909004839} {\bibfield  {journal} {\bibinfo  {journal}
  {International Journal of Quantum Information}\ }\textbf {\bibinfo {volume}
  {07}},\ \bibinfo {pages} {125--137} (\bibinfo {year} {2009})}\BibitemShut
  {NoStop}%
\bibitem [{\citenamefont {Pezz\`e}\ \emph {et~al.}(2018)\citenamefont
  {Pezz\`e}, \citenamefont {Smerzi}, \citenamefont {Oberthaler}, \citenamefont
  {Schmied},\ and\ \citenamefont {Treutlein}}]{Pezze2018}%
  \BibitemOpen
  \bibfield  {author} {\bibinfo {author} {\bibfnamefont {Luca}\ \bibnamefont
  {Pezz\`e}}, \bibinfo {author} {\bibfnamefont {Augusto}\ \bibnamefont
  {Smerzi}}, \bibinfo {author} {\bibfnamefont {Markus~K.}\ \bibnamefont
  {Oberthaler}}, \bibinfo {author} {\bibfnamefont {Roman}\ \bibnamefont
  {Schmied}}, \ and\ \bibinfo {author} {\bibfnamefont {Philipp}\ \bibnamefont
  {Treutlein}},\ }\bibfield  {title} {\enquote {\bibinfo {title} {Quantum
  metrology with nonclassical states of atomic ensembles},}\ }\href {\doibase
  10.1103/RevModPhys.90.035005} {\bibfield  {journal} {\bibinfo  {journal}
  {Rev. Mod. Phys.}\ }\textbf {\bibinfo {volume} {90}},\ \bibinfo {pages}
  {035005} (\bibinfo {year} {2018})}\BibitemShut {NoStop}%
\bibitem [{\citenamefont {Braun}\ \emph {et~al.}(2018)\citenamefont {Braun},
  \citenamefont {Adesso}, \citenamefont {Benatti}, \citenamefont {Floreanini},
  \citenamefont {Marzolino}, \citenamefont {Mitchell},\ and\ \citenamefont
  {Pirandola}}]{Braun2018}%
  \BibitemOpen
  \bibfield  {author} {\bibinfo {author} {\bibfnamefont {Daniel}\ \bibnamefont
  {Braun}}, \bibinfo {author} {\bibfnamefont {Gerardo}\ \bibnamefont {Adesso}},
  \bibinfo {author} {\bibfnamefont {Fabio}\ \bibnamefont {Benatti}}, \bibinfo
  {author} {\bibfnamefont {Roberto}\ \bibnamefont {Floreanini}}, \bibinfo
  {author} {\bibfnamefont {Ugo}\ \bibnamefont {Marzolino}}, \bibinfo {author}
  {\bibfnamefont {Morgan~W.}\ \bibnamefont {Mitchell}}, \ and\ \bibinfo
  {author} {\bibfnamefont {Stefano}\ \bibnamefont {Pirandola}},\ }\bibfield
  {title} {\enquote {\bibinfo {title} {Quantum-enhanced measurements without
  entanglement},}\ }\href {\doibase 10.1103/RevModPhys.90.035006} {\bibfield
  {journal} {\bibinfo  {journal} {Rev. Mod. Phys.}\ }\textbf {\bibinfo {volume}
  {90}},\ \bibinfo {pages} {035006} (\bibinfo {year} {2018})}\BibitemShut
  {NoStop}%
\bibitem [{\citenamefont {Sidhu}\ and\ \citenamefont {Kok}(2020)}]{Sidhu:2020}%
  \BibitemOpen
  \bibfield  {author} {\bibinfo {author} {\bibfnamefont {Jasminder~S.}\
  \bibnamefont {Sidhu}}\ and\ \bibinfo {author} {\bibfnamefont {Pieter}\
  \bibnamefont {Kok}},\ }\bibfield  {title} {\enquote {\bibinfo {title}
  {{Geometric perspective on quantum parameter estimation}},}\ }\href
  {https://doi.org/10.1116/1.5119961} {\bibfield  {journal} {\bibinfo
  {journal} {AVS Quantum Science}\ }\textbf {\bibinfo {volume} {2}} (\bibinfo
  {year} {2020})}\BibitemShut {NoStop}%
\bibitem [{\citenamefont {Gardiner}(2009)}]{Gardiner:2009}%
  \BibitemOpen
  \bibfield  {author} {\bibinfo {author} {\bibfnamefont {Crispin}\ \bibnamefont
  {Gardiner}},\ }\href {https://www.springer.com/gp/book/9783540707127} {\emph
  {\bibinfo {title} {Handbook of stochastic methods}}},\ \bibinfo {edition}
  {4th}\ ed.,\ \bibinfo {series} {Springer Series in Synergetics},
  Vol.~\bibinfo {volume} {13}\ (\bibinfo  {publisher} {Springer-Verlag Berlin
  Heidelberg},\ \bibinfo {year} {2009})\BibitemShut {NoStop}%
\bibitem [{\citenamefont {Spohn}(1976)}]{Spohn:1976}%
  \BibitemOpen
  \bibfield  {author} {\bibinfo {author} {\bibfnamefont {Herbert}\ \bibnamefont
  {Spohn}},\ }\bibfield  {title} {\enquote {\bibinfo {title} {Approach to
  equilibrium for completely positive dynamical semigroups of {$N$}-level
  systems},}\ }\href {\doibase https://doi.org/10.1016/0034-4877(76)90040-9}
  {\bibfield  {journal} {\bibinfo  {journal} {Reports on Mathematical Physics}\
  }\textbf {\bibinfo {volume} {10}},\ \bibinfo {pages} {189--194} (\bibinfo
  {year} {1976})}\BibitemShut {NoStop}%
\bibitem [{\citenamefont {Chru{\'s}ci{\'n}ski}(2022)}]{Chruscinski:2022}%
  \BibitemOpen
  \bibfield  {author} {\bibinfo {author} {\bibfnamefont {Dariusz}\ \bibnamefont
  {Chru{\'s}ci{\'n}ski}},\ }\bibfield  {title} {\enquote {\bibinfo {title}
  {{Dynamical maps beyond {M}arkovian regime}},}\ }\href {\doibase
  https://doi.org/10.1016/j.physrep.2022.09.003} {\bibfield  {journal}
  {\bibinfo  {journal} {Physics Reports}\ }\textbf {\bibinfo {volume} {992}},\
  \bibinfo {pages} {1--85} (\bibinfo {year} {2022})}\BibitemShut {NoStop}%
\bibitem [{\citenamefont {Gao}\ and\ \citenamefont {Han}(2012)}]{Gao:2012}%
  \BibitemOpen
  \bibfield  {author} {\bibinfo {author} {\bibfnamefont {Fuchang}\ \bibnamefont
  {Gao}}\ and\ \bibinfo {author} {\bibfnamefont {Lixing}\ \bibnamefont {Han}},\
  }\bibfield  {title} {\enquote {\bibinfo {title} {Implementing the
  {N}elder-{M}ead simplex algorithm with adaptive parameters},}\ }\href
  {\doibase https://doi.org/10.1007/s10589-010-9329-3} {\bibfield  {journal}
  {\bibinfo  {journal} {Computational Optimization and Applications}\ }\textbf
  {\bibinfo {volume} {262}},\ \bibinfo {pages} {259--277} (\bibinfo {year}
  {2012})}\BibitemShut {NoStop}%
\bibitem [{\citenamefont {Sachdev}(2011)}]{Sachdev:2011}%
  \BibitemOpen
  \bibfield  {author} {\bibinfo {author} {\bibfnamefont {Subir}\ \bibnamefont
  {Sachdev}},\ }\href {\doibase 10.1017/CBO9780511973765} {\emph {\bibinfo
  {title} {Quantum Phase Transitions}}}\ (\bibinfo  {publisher} {Cambridge
  University Press},\ \bibinfo {year} {2011})\BibitemShut {NoStop}%
\bibitem [{\citenamefont {Carollo}\ \emph {et~al.}(2020)\citenamefont
  {Carollo}, \citenamefont {Valenti},\ and\ \citenamefont
  {Spagnolo}}]{Carollo:2020}%
  \BibitemOpen
  \bibfield  {author} {\bibinfo {author} {\bibfnamefont {Angelo}\ \bibnamefont
  {Carollo}}, \bibinfo {author} {\bibfnamefont {Davide}\ \bibnamefont
  {Valenti}}, \ and\ \bibinfo {author} {\bibfnamefont {Bernardo}\ \bibnamefont
  {Spagnolo}},\ }\bibfield  {title} {\enquote {\bibinfo {title} {Geometry of
  quantum phase transitions},}\ }\href {\doibase
  https://doi.org/10.1016/j.physrep.2019.11.002} {\bibfield  {journal}
  {\bibinfo  {journal} {Physics Reports}\ }\textbf {\bibinfo {volume} {838}},\
  \bibinfo {pages} {1--72} (\bibinfo {year} {2020})}\BibitemShut {NoStop}%
\bibitem [{\citenamefont {Schultz}\ \emph {et~al.}(1964)\citenamefont
  {Schultz}, \citenamefont {Mattis},\ and\ \citenamefont {Lieb}}]{Schulz1964}%
  \BibitemOpen
  \bibfield  {author} {\bibinfo {author} {\bibfnamefont {T.~D.}\ \bibnamefont
  {Schultz}}, \bibinfo {author} {\bibfnamefont {D.~C.}\ \bibnamefont {Mattis}},
  \ and\ \bibinfo {author} {\bibfnamefont {E.~H.}\ \bibnamefont {Lieb}},\
  }\bibfield  {title} {\enquote {\bibinfo {title} {Two-dimensional {I}sing
  model as a soluble problem of many fermions},}\ }\href {\doibase
  10.1103/RevModPhys.36.856} {\bibfield  {journal} {\bibinfo  {journal} {Rev.
  Mod. Phys.}\ }\textbf {\bibinfo {volume} {36}},\ \bibinfo {pages} {856--871}
  (\bibinfo {year} {1964})}\BibitemShut {NoStop}%
\bibitem [{\citenamefont {Mbeng}\ \emph {et~al.}(2020)\citenamefont {Mbeng},
  \citenamefont {Russomanno},\ and\ \citenamefont {Santoro}}]{mbeng:2020}%
  \BibitemOpen
  \bibfield  {author} {\bibinfo {author} {\bibfnamefont {Glen~Bigan}\
  \bibnamefont {Mbeng}}, \bibinfo {author} {\bibfnamefont {Angelo}\
  \bibnamefont {Russomanno}}, \ and\ \bibinfo {author} {\bibfnamefont
  {Giuseppe~E.}\ \bibnamefont {Santoro}},\ }\bibfield  {title} {\enquote
  {\bibinfo {title} {The quantum {I}sing chain for beginners},}\ }\href
  {https://arxiv.org/abs/2009.09208} {\bibfield  {journal} {\bibinfo  {journal}
  {arXiv:2009.09208}\ } (\bibinfo {year} {2020})}\BibitemShut {NoStop}%
\bibitem [{\citenamefont {Rossini}\ and\ \citenamefont
  {Vicari}(2021)}]{Rossini:2021}%
  \BibitemOpen
  \bibfield  {author} {\bibinfo {author} {\bibfnamefont {Davide}\ \bibnamefont
  {Rossini}}\ and\ \bibinfo {author} {\bibfnamefont {Ettore}\ \bibnamefont
  {Vicari}},\ }\bibfield  {title} {\enquote {\bibinfo {title} {Coherent and
  dissipative dynamics at quantum phase transitions},}\ }\href {\doibase
  https://doi.org/10.1016/j.physrep.2021.08.003} {\bibfield  {journal}
  {\bibinfo  {journal} {Physics Reports}\ }\textbf {\bibinfo {volume} {936}},\
  \bibinfo {pages} {1--110} (\bibinfo {year} {2021})}\BibitemShut {NoStop}%
\bibitem [{\citenamefont {Weimer}\ \emph {et~al.}(2021)\citenamefont {Weimer},
  \citenamefont {Kshetrimayum},\ and\ \citenamefont {Or\'us}}]{Weimer:2021}%
  \BibitemOpen
  \bibfield  {author} {\bibinfo {author} {\bibfnamefont {Hendrik}\ \bibnamefont
  {Weimer}}, \bibinfo {author} {\bibfnamefont {Augustine}\ \bibnamefont
  {Kshetrimayum}}, \ and\ \bibinfo {author} {\bibfnamefont {Rom\'an}\
  \bibnamefont {Or\'us}},\ }\bibfield  {title} {\enquote {\bibinfo {title}
  {Simulation methods for open quantum many-body systems},}\ }\href {\doibase
  10.1103/RevModPhys.93.015008} {\bibfield  {journal} {\bibinfo  {journal}
  {Rev. Mod. Phys.}\ }\textbf {\bibinfo {volume} {93}},\ \bibinfo {pages}
  {015008} (\bibinfo {year} {2021})}\BibitemShut {NoStop}%
\bibitem [{\citenamefont {Demkowicz-Dobrza\'{n}ski}\ \emph
  {et~al.}(2020)\citenamefont {Demkowicz-Dobrza\'{n}ski}, \citenamefont
  {G\'{o}recki},\ and\ \citenamefont
  {Gu\c{t}\u{a}}}]{Demkowicz-Dobrzanski:2020}%
  \BibitemOpen
  \bibfield  {author} {\bibinfo {author} {\bibfnamefont {Rafa\l}\ \bibnamefont
  {Demkowicz-Dobrza\'{n}ski}}, \bibinfo {author} {\bibfnamefont {Wojciech}\
  \bibnamefont {G\'{o}recki}}, \ and\ \bibinfo {author} {\bibfnamefont
  {M\u{a}d\u{a}lin}\ \bibnamefont {Gu\c{t}\u{a}}},\ }\bibfield  {title}
  {\enquote {\bibinfo {title} {{Multi-parameter estimation beyond quantum
  Fisher information}},}\ }\href {\doibase 10.1088/1751-8121/ab8ef3} {\bibfield
   {journal} {\bibinfo  {journal} {Journal of Physics A: Mathematical and
  Theoretical}\ }\textbf {\bibinfo {volume} {53}},\ \bibinfo {pages} {363001}
  (\bibinfo {year} {2020})}\BibitemShut {NoStop}%
\bibitem [{\citenamefont {Albarelli}\ \emph {et~al.}(2020)\citenamefont
  {Albarelli}, \citenamefont {Barbieri}, \citenamefont {Genoni},\ and\
  \citenamefont {Gianani}}]{Albarelli:2020}%
  \BibitemOpen
  \bibfield  {author} {\bibinfo {author} {\bibfnamefont {F.}~\bibnamefont
  {Albarelli}}, \bibinfo {author} {\bibfnamefont {M.}~\bibnamefont {Barbieri}},
  \bibinfo {author} {\bibfnamefont {M.G.}\ \bibnamefont {Genoni}}, \ and\
  \bibinfo {author} {\bibfnamefont {I.}~\bibnamefont {Gianani}},\ }\bibfield
  {title} {\enquote {\bibinfo {title} {A perspective on multiparameter quantum
  metrology: From theoretical tools to applications in quantum imaging},}\
  }\href {\doibase https://doi.org/10.1016/j.physleta.2020.126311} {\bibfield
  {journal} {\bibinfo  {journal} {Physics Letters A}\ }\textbf {\bibinfo
  {volume} {384}},\ \bibinfo {pages} {126311} (\bibinfo {year}
  {2020})}\BibitemShut {NoStop}%
\bibitem [{\citenamefont {Turkington}(2013)}]{Turkington:2013}%
  \BibitemOpen
  \bibfield  {author} {\bibinfo {author} {\bibfnamefont {Darrell~A.}\
  \bibnamefont {Turkington}},\ }\href
  {https://doi.org/10.1017/CBO9781139424400} {\emph {\bibinfo {title}
  {Generalized Vectorization, Cross-Products, and Matrix Calculus}}}\ (\bibinfo
   {publisher} {Cambridge University Press},\ \bibinfo {year}
  {2013})\BibitemShut {NoStop}%
\bibitem [{\citenamefont {Manzano}(2020)}]{Manzano:2020}%
  \BibitemOpen
  \bibfield  {author} {\bibinfo {author} {\bibfnamefont {Daniel}\ \bibnamefont
  {Manzano}},\ }\bibfield  {title} {\enquote {\bibinfo {title} {{A short
  introduction to the {L}indblad master equation}},}\ }\href {\doibase
  10.1063/1.5115323} {\bibfield  {journal} {\bibinfo  {journal} {AIP Advances}\
  }\textbf {\bibinfo {volume} {10}},\ \bibinfo {pages} {025106} (\bibinfo
  {year} {2020})}\BibitemShut {NoStop}%
\bibitem [{\citenamefont {Landi}\ \emph {et~al.}(2022)\citenamefont {Landi},
  \citenamefont {Poletti},\ and\ \citenamefont {Schaller}}]{Landi_rev:2022}%
  \BibitemOpen
  \bibfield  {author} {\bibinfo {author} {\bibfnamefont {Gabriel~T.}\
  \bibnamefont {Landi}}, \bibinfo {author} {\bibfnamefont {Dario}\ \bibnamefont
  {Poletti}}, \ and\ \bibinfo {author} {\bibfnamefont {Gernot}\ \bibnamefont
  {Schaller}},\ }\bibfield  {title} {\enquote {\bibinfo {title} {Nonequilibrium
  boundary-driven quantum systems: Models, methods, and properties},}\ }\href
  {\doibase 10.1103/RevModPhys.94.045006} {\bibfield  {journal} {\bibinfo
  {journal} {Rev. Mod. Phys.}\ }\textbf {\bibinfo {volume} {94}},\ \bibinfo
  {pages} {045006} (\bibinfo {year} {2022})}\BibitemShut {NoStop}%
\bibitem [{\citenamefont {Campaioli}\ \emph {et~al.}(2024)\citenamefont
  {Campaioli}, \citenamefont {Cole},\ and\ \citenamefont
  {Hapuarachchi}}]{Campaioli:2024}%
  \BibitemOpen
  \bibfield  {author} {\bibinfo {author} {\bibfnamefont {Francesco}\
  \bibnamefont {Campaioli}}, \bibinfo {author} {\bibfnamefont {Jared~H.}\
  \bibnamefont {Cole}}, \ and\ \bibinfo {author} {\bibfnamefont {Harini}\
  \bibnamefont {Hapuarachchi}},\ }\bibfield  {title} {\enquote {\bibinfo
  {title} {Quantum master equations: Tips and tricks for quantum optics,
  quantum computing, and beyond},}\ }\href {\doibase
  10.1103/PRXQuantum.5.020202} {\bibfield  {journal} {\bibinfo  {journal} {PRX
  Quantum}\ }\textbf {\bibinfo {volume} {5}},\ \bibinfo {pages} {020202}
  (\bibinfo {year} {2024})}\BibitemShut {NoStop}%
\bibitem [{\citenamefont {Albert}\ and\ \citenamefont
  {Jiang}(2014)}]{Albert:2014}%
  \BibitemOpen
  \bibfield  {author} {\bibinfo {author} {\bibfnamefont {Victor~V.}\
  \bibnamefont {Albert}}\ and\ \bibinfo {author} {\bibfnamefont {Liang}\
  \bibnamefont {Jiang}},\ }\bibfield  {title} {\enquote {\bibinfo {title}
  {Symmetries and conserved quantities in {L}indblad master equations},}\
  }\href {\doibase 10.1103/PhysRevA.89.022118} {\bibfield  {journal} {\bibinfo
  {journal} {Phys. Rev. A}\ }\textbf {\bibinfo {volume} {89}},\ \bibinfo
  {pages} {022118} (\bibinfo {year} {2014})}\BibitemShut {NoStop}%
\bibitem [{\citenamefont {Albert}\ \emph {et~al.}(2016)\citenamefont {Albert},
  \citenamefont {Bradlyn}, \citenamefont {Fraas},\ and\ \citenamefont
  {Jiang}}]{Albert:2016}%
  \BibitemOpen
  \bibfield  {author} {\bibinfo {author} {\bibfnamefont {Victor~V.}\
  \bibnamefont {Albert}}, \bibinfo {author} {\bibfnamefont {Barry}\
  \bibnamefont {Bradlyn}}, \bibinfo {author} {\bibfnamefont {Martin}\
  \bibnamefont {Fraas}}, \ and\ \bibinfo {author} {\bibfnamefont {Liang}\
  \bibnamefont {Jiang}},\ }\bibfield  {title} {\enquote {\bibinfo {title}
  {Geometry and response of {L}indbladians},}\ }\href {\doibase
  10.1103/PhysRevX.6.041031} {\bibfield  {journal} {\bibinfo  {journal} {Phys.
  Rev. X}\ }\textbf {\bibinfo {volume} {6}},\ \bibinfo {pages} {041031}
  (\bibinfo {year} {2016})}\BibitemShut {NoStop}%
\bibitem [{\citenamefont {Mori}\ and\ \citenamefont
  {Shirai}(2020)}]{Mori:2020}%
  \BibitemOpen
  \bibfield  {author} {\bibinfo {author} {\bibfnamefont {Takashi}\ \bibnamefont
  {Mori}}\ and\ \bibinfo {author} {\bibfnamefont {Tatsuhiko}\ \bibnamefont
  {Shirai}},\ }\bibfield  {title} {\enquote {\bibinfo {title} {Resolving a
  discrepancy between {L}iouvillian gap and relaxation time in
  boundary-dissipated quantum many-body systems},}\ }\href {\doibase
  10.1103/PhysRevLett.125.230604} {\bibfield  {journal} {\bibinfo  {journal}
  {Phys. Rev. Lett.}\ }\textbf {\bibinfo {volume} {125}},\ \bibinfo {pages}
  {230604} (\bibinfo {year} {2020})}\BibitemShut {NoStop}%
\bibitem [{\citenamefont {Mori}\ and\ \citenamefont
  {Shirai}(2023)}]{Mori:2023}%
  \BibitemOpen
  \bibfield  {author} {\bibinfo {author} {\bibfnamefont {Takashi}\ \bibnamefont
  {Mori}}\ and\ \bibinfo {author} {\bibfnamefont {Tatsuhiko}\ \bibnamefont
  {Shirai}},\ }\bibfield  {title} {\enquote {\bibinfo {title} {Symmetrized
  {L}iouvillian {G}ap in {M}arkovian {O}pen {Q}uantum {S}ystems},}\ }\href
  {\doibase 10.1103/PhysRevLett.130.230404} {\bibfield  {journal} {\bibinfo
  {journal} {Phys. Rev. Lett.}\ }\textbf {\bibinfo {volume} {130}},\ \bibinfo
  {pages} {230404} (\bibinfo {year} {2023})}\BibitemShut {NoStop}%
\bibitem [{\citenamefont {Evans}(1977)}]{Evans:1977}%
  \BibitemOpen
  \bibfield  {author} {\bibinfo {author} {\bibfnamefont {David~E.}\
  \bibnamefont {Evans}},\ }\bibfield  {title} {\enquote {\bibinfo {title}
  {Irreducible quantum dynamical semigroups},}\ }\href {\doibase
  10.1007/BF01614091} {\bibfield  {journal} {\bibinfo  {journal} {Commun. Math.
  Phys.}\ }\textbf {\bibinfo {volume} {54}},\ \bibinfo {pages} {293--297}
  (\bibinfo {year} {1977})}\BibitemShut {NoStop}%
\bibitem [{\citenamefont {Nigro}(2019)}]{Nigro:2019}%
  \BibitemOpen
  \bibfield  {author} {\bibinfo {author} {\bibfnamefont {Davide}\ \bibnamefont
  {Nigro}},\ }\bibfield  {title} {\enquote {\bibinfo {title} {On the uniqueness
  of the steady-state solution of the
  {L}indblad-{G}orini-{K}ossakowski-{S}udarshan equation},}\ }\href {\doibase
  10.1088/1742-5468/ab0c1c} {\bibfield  {journal} {\bibinfo  {journal} {Journal
  of Statistical Mechanics: Theory and Experiment}\ }\textbf {\bibinfo {volume}
  {2019}},\ \bibinfo {pages} {043202} (\bibinfo {year} {2019})}\BibitemShut
  {NoStop}%
\bibitem [{\citenamefont {Zhang}\ and\ \citenamefont
  {Barthel}(2024)}]{Zhang:2024}%
  \BibitemOpen
  \bibfield  {author} {\bibinfo {author} {\bibfnamefont {Yikang}\ \bibnamefont
  {Zhang}}\ and\ \bibinfo {author} {\bibfnamefont {Thomas}\ \bibnamefont
  {Barthel}},\ }\bibfield  {title} {\enquote {\bibinfo {title} {Criteria for
  {D}avies irreducibility of {M}arkovian quantum dynamics},}\ }\href {\doibase
  10.1088/1751-8121/ad2a1e} {\bibfield  {journal} {\bibinfo  {journal} {Journal
  of Physics A: Mathematical and Theoretical}\ }\textbf {\bibinfo {volume}
  {57}},\ \bibinfo {pages} {115301} (\bibinfo {year} {2024})}\BibitemShut
  {NoStop}%
\end{thebibliography}%

\end{document}